\documentclass[aps,prl,twocolumn,superscriptaddress]{revtex4-1}

\usepackage{graphicx}
\usepackage{dcolumn}
\usepackage{bm}
\usepackage{bbm}
\usepackage{dsfont}
\usepackage{braket}
\usepackage{amsmath}
\usepackage{amssymb}
\usepackage{amsfonts}
\usepackage{xcolor}
\usepackage[colorlinks=true,citecolor=blue,linkcolor=blue,urlcolor=blue]{hyperref}
\usepackage[utf8]{inputenc}
\usepackage[T1]{fontenc}
\usepackage{textcomp}

\renewcommand{\Re}{\operatorname{Re}}

\newcommand{\iu}{\mathrm{i}}
\newcommand{\eu}{\mathrm{e}}

\newcommand{\Beta}{\mathrm{B}}

\newcommand{\umat}[1]{{\underline{\underline{#1}}}}
\newcommand{\uvec}[1]{{\underline{#1}}}

\newcommand{\beginsupplement}{%
        \setcounter{table}{0}
        \renewcommand{\thetable}{S\arabic{table}}%
        \setcounter{figure}{0}
        \renewcommand{\thefigure}{S\arabic{figure}}%
        \setcounter{equation}{0}
        \renewcommand{\theequation}{S\arabic{equation}}%
        \setcounter{secnumdepth}{4}
        \renewcommand{\thesection}{\Roman{section}}
     }

\begin{document}

\title{Distinct Optical Excitation Mechanisms of a Coherent Magnon \\ in a van der Waals Antiferromagnet}

\author{Clifford J. Allington}
\altaffiliation{These authors contributed equally to this work.}
\affiliation{Department of Physics, Massachusetts Institute of Technology, Cambridge, Massachusetts 02139, USA}

\author{Carina A. Belvin}
\altaffiliation{These authors contributed equally to this work.}
\affiliation{Department of Physics, Massachusetts Institute of Technology, Cambridge, Massachusetts 02139, USA}

\author{Urban F. P. Seifert}
\affiliation{Kavli Institute for Theoretical Physics, University of California, Santa Barbara, California 93106, USA}

\author{Mengxing Ye}
\affiliation{Kavli Institute for Theoretical Physics, University of California, Santa Barbara, California 93106, USA}
\affiliation{Department of Physics and Astronomy, University of Utah, Salt Lake City, Utah 84112, USA}

\author{Tommy Tai}
\affiliation{Department of Physics, Massachusetts Institute of Technology, Cambridge, Massachusetts 02139, USA}

\author{Edoardo Baldini}
\affiliation{Department of Physics, Massachusetts Institute of Technology, Cambridge, Massachusetts 02139, USA}

\author{Suhan Son}
\affiliation{Center for Quantum Materials, Department of Physics and Astronomy, Seoul National University, Seoul 08826, Korea}

\author{Junghyun Kim}
\affiliation{Center for Quantum Materials, Department of Physics and Astronomy, Seoul National University, Seoul 08826, Korea}

\author{Jaena Park}
\affiliation{Center for Quantum Materials, Department of Physics and Astronomy, Seoul National University, Seoul 08826, Korea}

\author{Je-Geun Park}
\affiliation{Center for Quantum Materials, Department of Physics and Astronomy, Seoul National University, Seoul 08826, Korea}

\author{Leon Balents}
\affiliation{Kavli Institute for Theoretical Physics, University of California, Santa Barbara, California 93106, USA}

\author{Nuh Gedik}
\altaffiliation{Email: gedik@mit.edu}
\affiliation{Department of Physics, Massachusetts Institute of Technology, Cambridge, Massachusetts 02139, USA}

\date{February 10, 2025}

\begin{abstract}
The control of antiferromagnets with ultrashort optical pulses has emerged as a prominent field of research. Tailored laser excitation can launch coherent spin waves at terahertz frequencies, yet a comprehensive description of their generation mechanisms is still lacking despite extensive efforts. Using terahertz emission spectroscopy, we investigate the generation of a coherent magnon mode in the van der Waals antiferromagnet NiPS$_3$ under a range of photoexcitation conditions. By tuning the pump photon energy from transparency to resonant with a $d$-$d$ transition, we reveal a striking change in the coherent magnon's dependence on the pump polarization, indicating two distinct excitation mechanisms. Our findings provide a strategy for the manipulation of magnetic modes via photoexcitation around sub-gap electronic states.
\end{abstract}

\maketitle

Antiferromagnets are desirable for the development of innovative spin-based devices due to their lack of stray fields, insensitivity to external magnetic perturbations, and intrinsically fast spin dynamics \cite{jungwirth2016antiferromagnetic,baltz2018antiferromagnetic,nvemec2018antiferromagnetic}. These advantages also pose challenges to the efficient detection and manipulation of their magnetic order. In recent years, there has been substantial progress in using ultrashort laser pulses to control spin dynamics in antiferromagnets \cite{kimel2005ultrafast,kalashnikova2007impulsive,kimel2009inertia,kirilyuk2010ultrafast,nishitani2010terahertz,kampfrath2011coherent,ivanov2014spin,kubacka2014large,mikhaylovskiy2015ultrafast,baierl2016nonlinear,nova2017effective,schlauderer2019temporal,disa2020polarizing,afanasiev2021ultrafast,hortensius2021coherent}. These studies have revealed various thermal and non-thermal microscopic mechanisms for the light-induced generation of coherent spin waves. Non-thermal methods are especially attractive because they avoid unwanted heating and enable ultrafast control.

Canonical examples are the inverse Faraday effect \cite{kimel2005ultrafast,satoh2010spin} and the inverse Cotton-Mouton effect \cite{kalashnikova2007impulsive,tzschaschel2017ultrafast} for photoexcitation with circularly and linearly polarized light, respectively. These magneto-optical effects, which have been described phenomenologically as impulsive stimulated Raman scattering (ISRS) \cite{gridnev2008phenomenological,kirilyuk2010ultrafast}, can be observed when the photon energy of the laser pulse lies in a region of optical transparency in the material. Microscopic mechanisms responsible for these effects include light-induced Zeeman splittings and magnetocrystalline anisotropies by coupling to virtual orbital $d$-$d$ transitions \cite{hansteen2005femtosecond,seifert2022ultrafast}, virtual doublon excitations \cite{nishitani2012coherent}, or a direct coupling of the driving light to (interacting) magnon excitations \cite{seifert2019optical}. A characteristic signature of these effects is a strong dependence of the magnon amplitude on the driving light's polarization. On the other hand, if the system is driven in a regime of weak absorption, the pump laser can resonantly couple to spin-forbidden $d$-$d$ transitions \cite{mikhaylovskiy2020resonant}, to intermediate excitonic states \cite{belvin2021exciton,bossini2021ultrafast}, or to phonon excitations, which in turn couple to magnons \cite{nova2017effective,afanasiev2021ultrafast}. In these cases, the magnon amplitude is typically found to be independent of the pump polarization. However, it is unknown whether there exists a connection between the mechanisms in the regimes of transparency and weak absorption since no comprehensive theoretical framework for coherent magnon generation has been put forth to date. This is in contrast to other types of collective modes such as coherent phonons, whose excitation is described by the two Raman tensor theory \cite{stevens2002coherent}. To this end, a thorough experimental characterization of the evolution of a coherent magnon mode across these two regions of photoexcitation would provide vital information about the underlying microscopic processes.

Here, we track a coherent magnon under tunable photoexcitation from transparency to weak absorption using terahertz (THz) emission spectroscopy. As a model system, we choose the van der Waals antiferromagnet NiPS$_3$, which exhibits a variety of sub-gap optical features \cite{piacentini1982optical,banda1986optical,grasso1986optical,joy1992optical,kang2020coherent,hwangbo2021highly,wang2021spin,wang2022electronic,toyoda2024phase} and a long-wavelength magnon mode around 1.3~THz \cite{belvin2021exciton}. By varying the optical excitation energy from a region of transparency to resonant with an orbital excitation at 1.1~eV, we observe the launch of this magnon mode coherently and map its dependence on the pump polarization to elucidate its generation mechanisms. We find that the coherent spin wave behavior in the two photoexcitation limits reveals markedly different characteristics, demonstrating the existence of two distinct mechanisms of excitation. In transparency, we find that the coherent magnon generation can be explained by magneto-optical effects, yielding a characteristic polarization dependence. In contrast, when driving on resonance, the system can absorb energy through transitions to excited levels, and its relaxation back to the ground-state manifold occurs via dissipative processes. When these dissipative processes are constrained to the global symmetry of the system, rather than the local symmetry of the Ni$^{2+}$ ion, then no characteristic polarization dependence is expected for the generation of the coherent magnon.

\begin{figure}[t]
\begin{center}
\includegraphics[width=\columnwidth]{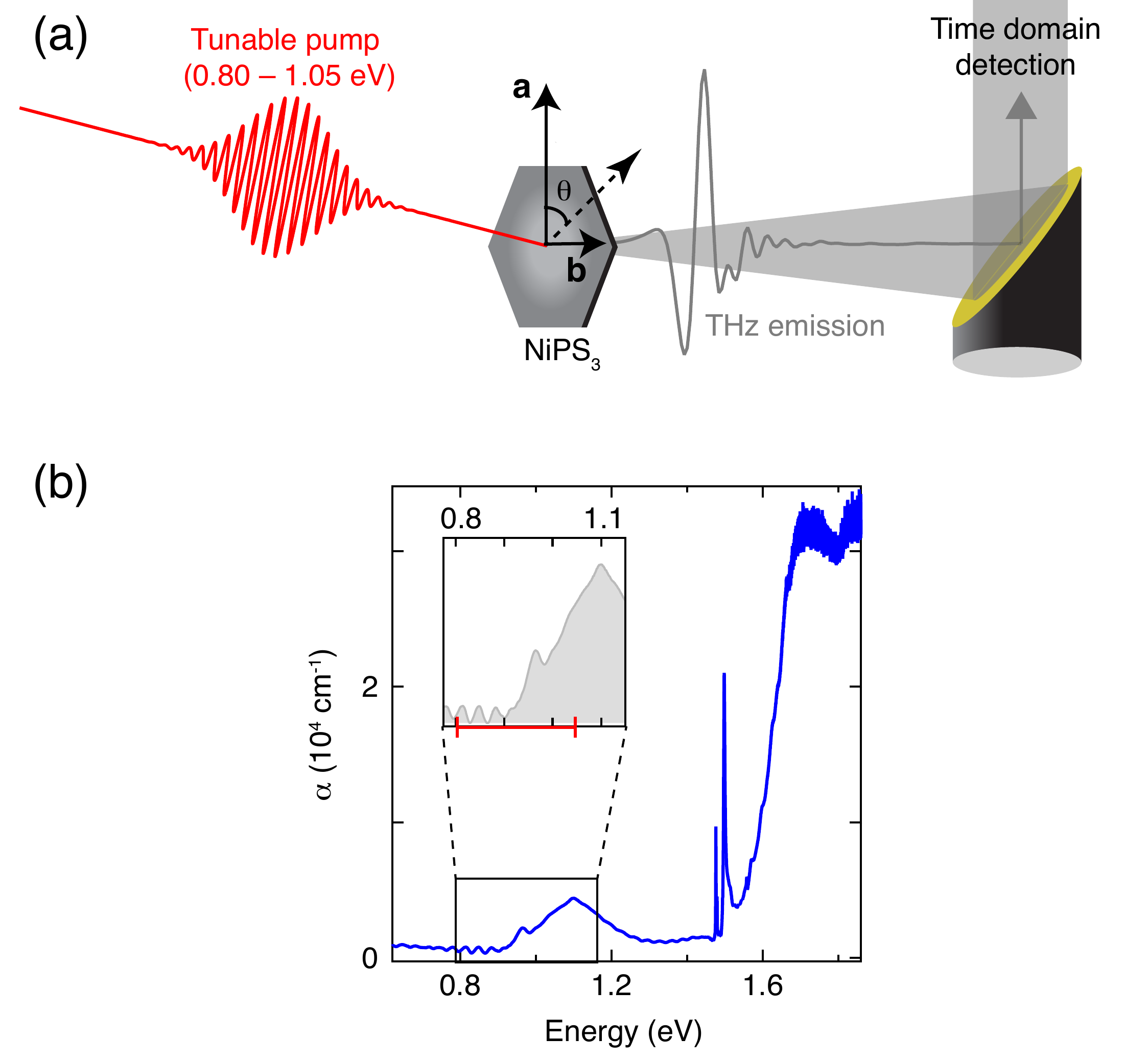}
\caption{(a) Depiction of our THz emission spectroscopy setup. The pump pulse is tunable in the near infrared (0.80--1.05~eV) and is linearly polarized with an angle $\theta$ with respect to the crystallographic $a$-axis of NiPS$_3$. The emitted THz radiation from the sample is detected in the time domain using electro-optic sampling. (b) Optical absorption spectrum of NiPS$_3$ at 20~K (the data is from Ref.~\cite{belvin2021exciton}). The inset shows the on-site $d$-$d$ transition around 1.1~eV (which is split due to a trigonal distortion) and the transparent region below this energy.}
\label{fig:Fig1}
\end{center}
\end{figure}

Our ultrafast THz emission spectroscopy setup is depicted in Fig.~\ref{fig:Fig1}(a). A tunable pump pulse ($\sim70$~fs) in the near-infrared range (0.80--1.05~eV) is incident on our NiPS$_3$ sample with a linear polarization angle $\theta$ relative to the crystallographic $a$-axis. As a function of time delay after the pump pulse arrival, we detect the THz radiation emitted by the sample via electro-optic sampling. This technique can directly measure in the time domain the radiation from coherent magnons, produced when the oscillating dipoles are oriented perpendicular to the light propagation direction \cite{nishitani2010terahertz,nishitani2012coherent,mikhaylovskiy2015ultrafast}.

The optical absorption spectrum of NiPS$_3$ [Fig.~\ref{fig:Fig1}(b)] contains several features including on-site $d$-$d$ transitions between electronic states of the Ni$^{2+}$ ion that are split by its octahedral coordination. We focus on the broad peak around 1.1~eV (inset) corresponding to the transition $^3A_{2g} \rightarrow\, ^3T_{2g}$, which is spin allowed ($\Delta S=0$). This transition is split due to a trigonal distortion of the octahedral crystal field, giving rise to a small shoulder around 0.95~eV on the left side of the main broad peak centered around 1.1~eV \cite{piacentini1982optical,banda1986optical,grasso1986optical}. By tuning the pump photon energy throughout the range indicated in red in the inset of Fig.~\ref{fig:Fig1}(b), we can excite the system in transparency or in resonance with this $d$-$d$ transition. We note that our maximum pump photon energy is 1.05~eV, which does not allow us to pump above the $d$-$d$ transition. All measurements were performed at 15~K, well below the N\'eel temperature of NiPS$_3$ ($T_N=157$~K).

\begin{figure*}
\begin{center}
\includegraphics[width=2\columnwidth]{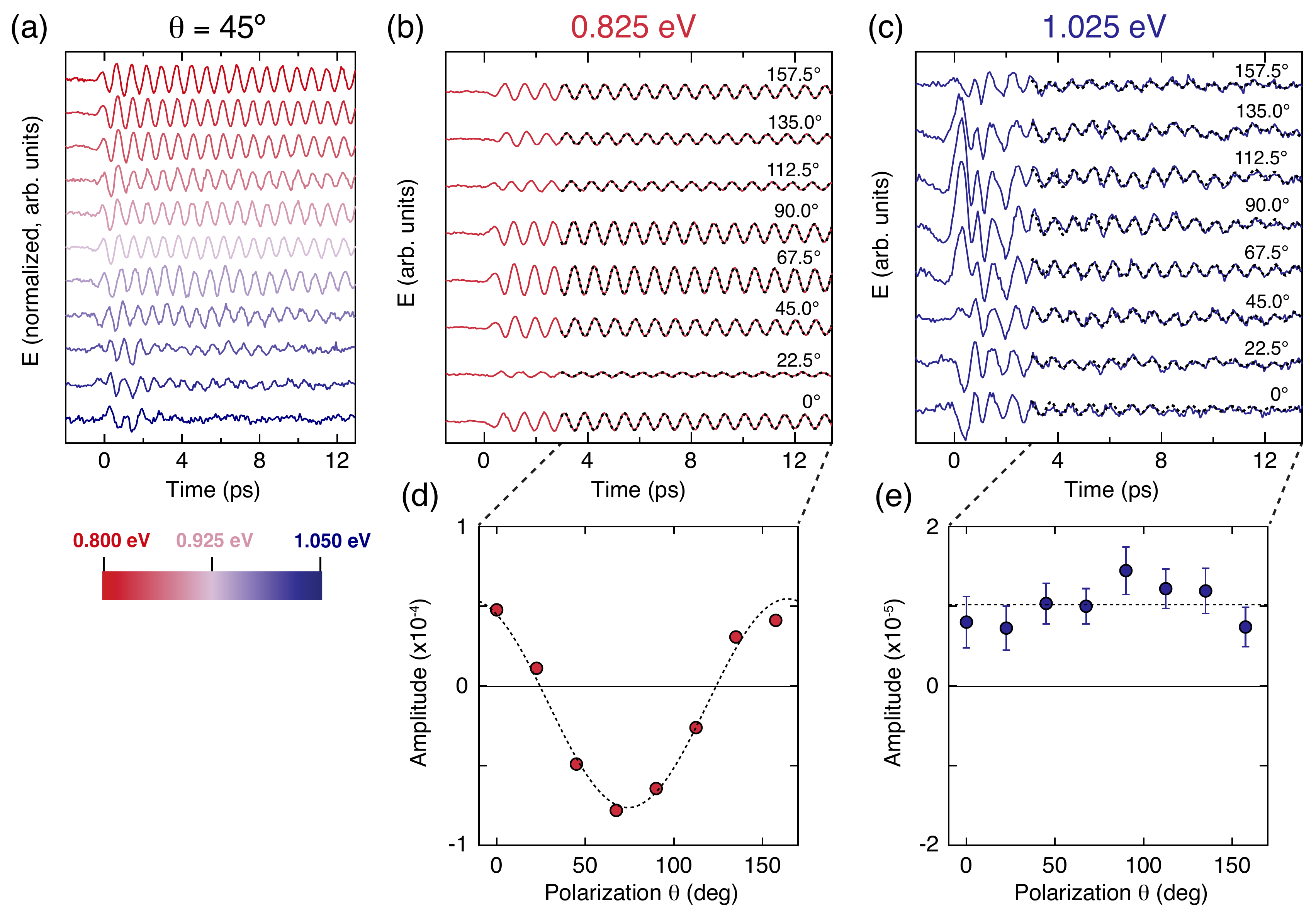}
\caption{(a) THz emission signal at 15~K as a function of time after the pump pulse arrival for a range of pump photon energies. The pump polarization is fixed at $\theta=45^\circ$. The curves are normalized and displaced vertically for clarity. (b), (c) Dependence of the THz emission signal on the pump polarization for pump photon energies 0.825 eV (in the transparent regime) and 1.025 eV (resonant with the $d$-$d$ transition), respectively. Fits of each curve to sinusoidal oscillations are shown in black dashed lines. The pump polarization angle $\theta$ is indicated for each curve on the right side of the plots. The traces in (b) and (c) are vertically offset for clarity. (d), (e) Amplitude of the magnon oscillation extracted from fits to the raw data in (b) and (c), respectively. The error bars represent the 95\% confidence interval from the fits (in (d), the size of the points is larger than the error bars). For the transparent region (d), we observe a cos($2\theta$) behavior (black dashed line) of the magnon amplitude, whereas for the resonant case (e), the magnon amplitude is nearly independent of the pump polarization angle.}
\label{fig:Fig2}
\end{center}
\end{figure*}

The THz emission data is presented in Fig.~\ref{fig:Fig2}(a). At all pump photon energies, we observe oscillations as a function of time after the pump pulse with a frequency around 1.3~THz. This frequency corresponds to that of the zone-center magnon mode detected previously in the equilibrium THz absorption \cite{belvin2021exciton}. Moreover, the frequency of the THz emission oscillations softens as a function of temperature with an order parameter behavior and disappears above $T_N=157$~K (see Fig.~S8 in the Supplemental Material). Therefore, we indeed observe the same magnon mode for all pump photon energies, enabling us to study the mechanisms by which it is coherently generated as the pump is tuned below or in resonance with the $d$-$d$ transition. The initial transient that precedes the magnon oscillations in the resonant case is present at all temperatures and is likely due to surface optical rectification (see Fig.~S5 in the Supplemental Material).

In order to determine the mechanisms launching the coherent magnon in these two regimes, we examine the dependence of the THz emission on the linear polarization angle of the pump pulse. Figures~\ref{fig:Fig2}(b) and~\ref{fig:Fig2}(c) display the pump polarization dependence of the THz emission signal for pump photon energies 0.825~eV (in transparency) and 1.025~eV (resonant with the $d$-$d$ transition), respectively. We fit each of the raw data curves with damped sinusoids (starting at a time after the initial optical rectification signal) and extract the magnon amplitude, shown in Figs.~\ref{fig:Fig2}(d) and~\ref{fig:Fig2}(e). Only one sinusoid is needed in transparency, whereas two sinusoids are used on resonance to also account for a lower-energy magnon as described below. There is a striking difference in the polarization dependences for these two pump photon energies. In transparency, the magnon amplitude follows a cos($2\theta$) behavior, which is indicative of a (non-thermal) magneto-optic effect \cite{kalashnikova2007impulsive,tzschaschel2017ultrafast}. There is also a small isotropic offset similar to what is observed in Ref.~\cite{tzschaschel2017ultrafast}. On the other hand, in the resonant regime, the magnon amplitude is nearly independent of the pump polarization angle and the amplitude notably does not cross zero at any point. 

To further understand the mechanisms of coherent magnon generation in these two photoexcitation limits, we examine the oscillation amplitude as a function of absorbed pump fluence. In transparency the amplitude of the magnon mode grows linearly with pump fluence, while for the resonant case a saturation occurs at high fluences (see Figs.~S6 and S7 in the Supplemental Material). 

The markedly different polarization and fluence dependences in these two photoexcitation regimes indicate two distinct microscopic mechanisms for coherent magnon generation. In the case of transparency, the linear dependence of the magnon amplitude on pump fluence and its cos($2\theta$) polarization dependence are clear signatures of magneto-optic effects previously studied \cite{seifert2022ultrafast,afanasiev2021controlling}. A recent theoretical work on NiPS$_3$ developed a microscopic picture of this effect using a light-induced Floquet spin Hamiltonian acting on the Ni$^{2+}$ ion in a perturbative framework \cite{seifert2022ultrafast}. This work predicted a pump polarization dependence of the magnon amplitude which matches the experimentally observed behavior in Fig.~\ref{fig:Fig2}(d). It also presented a theory for magnon excitation by driving an orbital $d$-$d$ transition near but not on resonance. Such a mechanism, which can be understood as an ultrafast change in the magnetic anisotropy \cite{hansteen2005femtosecond}, was shown to coherently generate a lower-energy magnon mode at 0.3~THz and explain its cos($2\theta$) pump polarization dependence as observed in a previous experimental study using pump-probe Faraday rotation \cite{afanasiev2021controlling}. We also observe this 0.3~THz mode and the same variation with pump polarization, though its signal is much  weaker in our (slightly oblique incidence) THz emission experiment compared with Faraday rotation due to geometrical considerations of in-plane vs. out-of-plane magnetization oscillations (see Sec.~IV and Fig.~S4 in the Supplemental Material). The 1.3~THz magnon mode, however, which is the main focus of this work, displays instead an amplitude (mostly) independent of the pump polarization on resonance, suggesting that additional theoretical considerations are needed to describe its coherent generation. 

\begin{figure}[t]
\begin{center}
\includegraphics[width=\columnwidth]{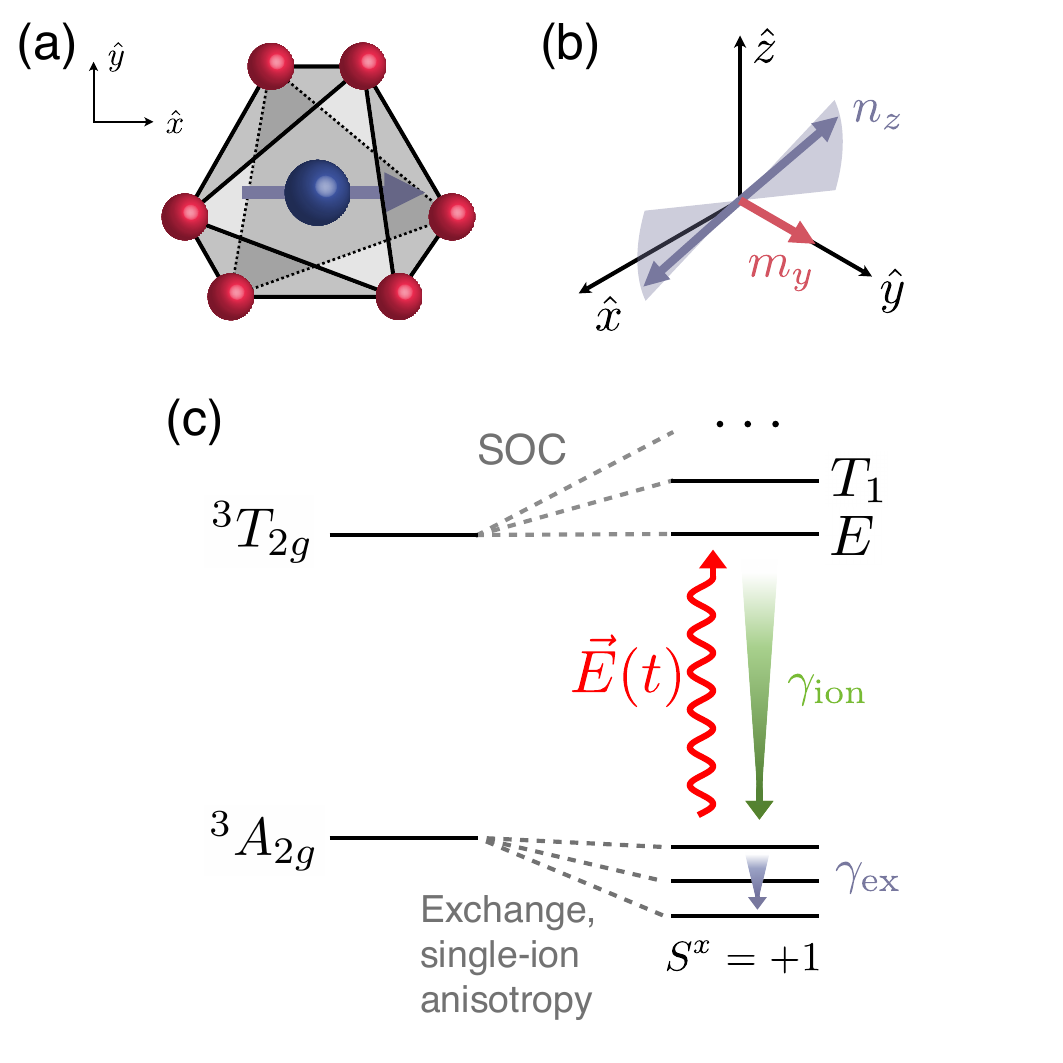}
\caption{(a) Magnetic Ni$^{2+}$ ion (blue) in an octahedral crystal field due to the surrounding S atoms. In equilibrium, the exchange interactions and single-ion anisotropies lead to a zigzag antiferromagnetic order with moments pointing along the $\pm \hat{x}$-axis. (b) Illustration of the spin precession of the 1.3~THz magnon mode. The N\'eel vector $n$ oscillates in the $xz$-plane, inducing a finite time-dependent uniform magnetization $m$ along $\hat{y}$ as a conjugate variable. (c) Schematic illustration of the dissipation-controlled magnon excitation process. Under spin-orbit coupling (SOC), the $^3T_{2g}$ orbital multiplet gives rise (among others) to $E$ and $T_1$ multiplet states, while the $S=1$ ground-state manifold in equilibrium is weakly split by single-ion anisotropies and exchange fields. Driving the system with light $\vec{E}(t)$ in resonance with the $d$-$d$ transition leads to a population of the excited levels. These states decay with a rate $\gamma_\text{ion}$ to the $S=1$ manifold. Within this manifold, there is an additional dissipative process with decay rate $\gamma_{\mathrm{ex}}$ that tends to align the moments with the exchange field of the respective Ni$^{2+}$ ion. The steady-state solution (accounting for resonant driving and both dissipation processes) has a finite staggered magnetization along $\hat{z}$ (i.e. $n^z\neq 0$), thus setting the initial conditions for the relaxation after the pump, according to the semiclassical equations of motion of the magnon mode.}
\label{fig:Fig3}
\end{center}
\end{figure}

To capture this different behavior for the 1.3~THz mode under resonant photoexcitation, we must go beyond second-order perturbation theory \cite{seifert2022ultrafast}, which breaks down when the detuning energy is smaller than the dissipation rate. That is, when the incident pump pulse is strongly resonant with the $d$-$d$ transition, dissipation becomes significant such that we must model the Ni$^{2+}$ ion within a Lindbladian picture, where it undergoes both unitary time evolution (induced by driving with the pump electric field) as well as dissipative dynamics due to the spontaneous decay from the excited state back to the ground-state manifold. Therefore, the state of the Ni$^{2+}$ ion, given by the density matrix $\rho$, evolves as an open quantum system according to the Lindblad equation
\begin{equation}
	\partial_t \rho = \iu [\rho,H] + \sum_j \gamma_j \left[L_j \rho L_j^\dagger - \frac{1}{2} \{L_j^\dagger L_j,\rho \} \right],
\end{equation}
where the $L_j$ are jump operators for spontaneous transitions, and the $\gamma_j \geq 0$ are decay rates (see Sec.~IX in the Supplemental Material for details of the theoretical model). This decay process is illustrated schematically in Fig.~\ref{fig:Fig3}(c). In essence, this process provides the initial conditions to launch the coherent spin precession of the 1.3~THz magnon mode (depicted in Fig.~\ref{fig:Fig3}(b)). The decay from the excited state to the $S=1$ manifold, as well as the decay within the $S=1$ manifold, all occur within the duration of the pump pulse. The second decay process (within the $S=1$ manifold) results in a finite staggered magnetization along the $z$-direction, which is equal but opposite for the two magnetic sublattices, and this initiates the spin precession under a semi-classical equation of motion. 

As a function of the pump electric field, this microscopic picture predicts an initial rise of the magnon amplitude followed by a plateau (see Fig.~S11 in the Supplemental Material). Such behavior is consistent with the increase and then saturation that we observe for the magnon amplitude with our experimentally accessible pump fluence values (see Figs.~S6 and S7 in the Supplemental Material). This theory also accounts for the (mostly) pump polarization independent response of the magnon amplitude (see Fig.~S13 in the Supplemental Material) that we observe in Fig.~\ref{fig:Fig2}(e). Therefore, our experimental findings demonstrate a new type of coherent magnon generation mechanism that stems from the interplay of resonant driving and non-unitary evolution of a system subject to dissipation. Further evidence for this dissipative mechanism is found in the measured pump-induced THz transmission of NiPS$_3$ in the same photoexcitation region (see Fig.~S9 in the Supplemental Material), which reveals the creation of a small amount of charge carriers only for resonant pump photon energies. This observation is consistent with a previous study on a different antiferromagnetic material that demonstrated a lack of heating of electrons solely in regions of nearly zero absorption \cite{bossini2014controlling}. 

We note that this mechanism including dissipation can also be used to describe the coherent generation of the 0.3~THz mode. Since this lower-energy mode is less sensitive to dissipation, the earlier modeling in Ref.~\cite{seifert2022ultrafast} was sufficient to account for its experimentally observed trends \cite{afanasiev2021controlling}. Additionally, it can be shown that the particular symmetry of each of the two modes gives rise to their different dependences on the pump polarization on resonance. In particular, the 0.3~THz magnon mode, which corresponds to staggered fluctuations along the $y$-axis, cannot be excited at certain polarization angles of the pump beam, i.e. along the $x$- and $y$-axes. When the pump polarization is along either of these axes, the vertical mirror (in-plane two-fold rotation followed by a lattice translation) remains a good symmetry in the zigzag ordered system. As the 0.3~THz magnon mode breaks both symmetries, it is forbidden to be excited when the pump polarization is along $x$ or $y$. However, the 1.3~THz magnon mode is not forbidden on symmetry grounds. Lastly, we remark that the coherent magnon generation mechanisms demonstrated in this work for the 1.3~THz mode are distinct from the mechanism reported in Ref.~\cite{belvin2021exciton}, in which an exciton coupled to the antiferromagnetic order is responsible for launching this coherent magnon (notably seen in the redshift of the magnon energy with increasing pump fluence that is proportional to the density of photogenerated excitons). In addition, our results also differ from those of Ref.~\cite{mikhaylovskiy2020resonant}, which demonstrates that photoexciting a spin-forbidden $d$-$d$ transition (involving a spin flip from the ground state to the excited states) causes an ultrafast change in the exchange interactions in iron oxides with canted antiferromagnetism.

In summary, we demonstrate two distinct optical excitation mechanisms of the same coherent magnon mode when a laser pulse is tuned from transparency to resonant with a $d$-$d$ transition in NiPS$_3$. Since $d$-$d$ excitations and neighboring regions of optical transparency are ubiquitous in insulating transition metal compounds, these concepts can be extended to a broad class of magnetic materials, including two-dimensional magnets. Our findings offer a protocol for the selective control over the properties of coherent magnon generation by tuning across sub-gap electronic resonances. At the same time, our results highlight the intricate microscopic processes responsible for the photoexcitation of coherent magnons. We reveal a new mechanism based on dissipation under resonant conditions, adding to the breadth of generation mechanisms needed to describe coherent magnons far beyond the two Raman tensor framework for coherent phonons.\\

We acknowledge useful discussions with Karna Morey, Bryan Fichera, and Roberto Merlin. Work at MIT was supported by the US Department of Energy, BES DMSE (data taking and analysis), and by the Gordon and Betty Moore Foundation's EPiQS Initiative grant GBMF9459 (instrumentation). The work at SNU was supported by the Leading Researcher Program of Korea's National Research Foundation (Grant No. 2020R1A3B2079375). C.A.B. and C.J.A. acknowledge additional support from the National Science Foundation Graduate Research Fellowship under Grant No. 1745302 and the Department of Defense (DoD) through the National Defense Science \& Engineering Graduate (NDSEG) Fellowship Program, respectively. T.T. acknowledges support from the National Science Scholarship under A*STAR of Singapore. E.B. acknowledges additional support from the Swiss National Science Foundation under fellowships P2ELP2-172290 and P400P2-183842. U.F.P.S. was supported by the Deutsche Forschungs-gemeinschaft (DFG, German Research Foundation) through a Walter Benjamin fellowship, Project ID No. 449890867 and the DOE office of BES, through award number DE-SC0020305. L.B. was supported by the NSF CMMT program under Grant No. DMR-2116515, the Gordon and Betty Moore Foundation through Grant GBMF8690, and by the Simons Collaboration on Ultra-Quantum Matter, which is a grant from the Simons Foundation (651440). M.Y. was supported by the Gordon and Betty Moore Foundation through Grant GBMF8690 to UCSB, by a grant from the Simons Foundation (216179, LB), and by the National Science Foundation under Grant No. NSF PHY-1748958.

\clearpage
\beginsupplement
\onecolumngrid
\begin{center}
\textbf{\large Supplemental Material for ``Distinct Optical Excitation Mechanisms of a Coherent Magnon in a van der Waals Antiferromagnet''}
\end{center}\hfill\break
\twocolumngrid

\section{Experimental Details}
\label{sec:expdet}

High-quality single crystals of NiPS$_3$ were grown using the chemical vapor transport method. The details of the growth conditions are described in a previous work \cite{kuo2016exfoliation}. The crystals were characterized using in-plane magnetic susceptibility and heat capacity measurements to reveal that $T_N \sim 157$ K (the data is shown in Ref.~\cite{belvin2021exciton}). In order to determine the orientation of the crystal axes for our measurements, we performed Laue X-ray diffraction.

The ultrafast terahertz (THz) emission spectroscopy measurements were carried out using a Ti:Sapphire amplified laser system with 35~fs pulses at a central photon energy of 1.55~eV and a repetition rate of 1 kHz. To obtain a tunable source for our pump pulses, the fundamental of the laser was sent into an optical parametric amplifier (OPA), which outputs pulses in the near infrared with approximately $70$~fs pulse duration and photon energies ranging from 0.80 to 1.05~eV. The pump pulses were incident on the sample at an angle of $\sim15^{\circ}$ away from the normal direction and the pump spot size was $\sim4$ mm in diameter. The sample thickness was 750 $\mu$m. The THz emission from the sample was detected in the time domain using electro-optic sampling in a ZnTe crystal with a 1.55~eV gate pulse. The sample was held in a vacuum cryostat that allows for temperatures from 4 to 300~K.

\section{Optical Absorption}

\begin{figure}
\begin{center}
\includegraphics[width=\columnwidth]{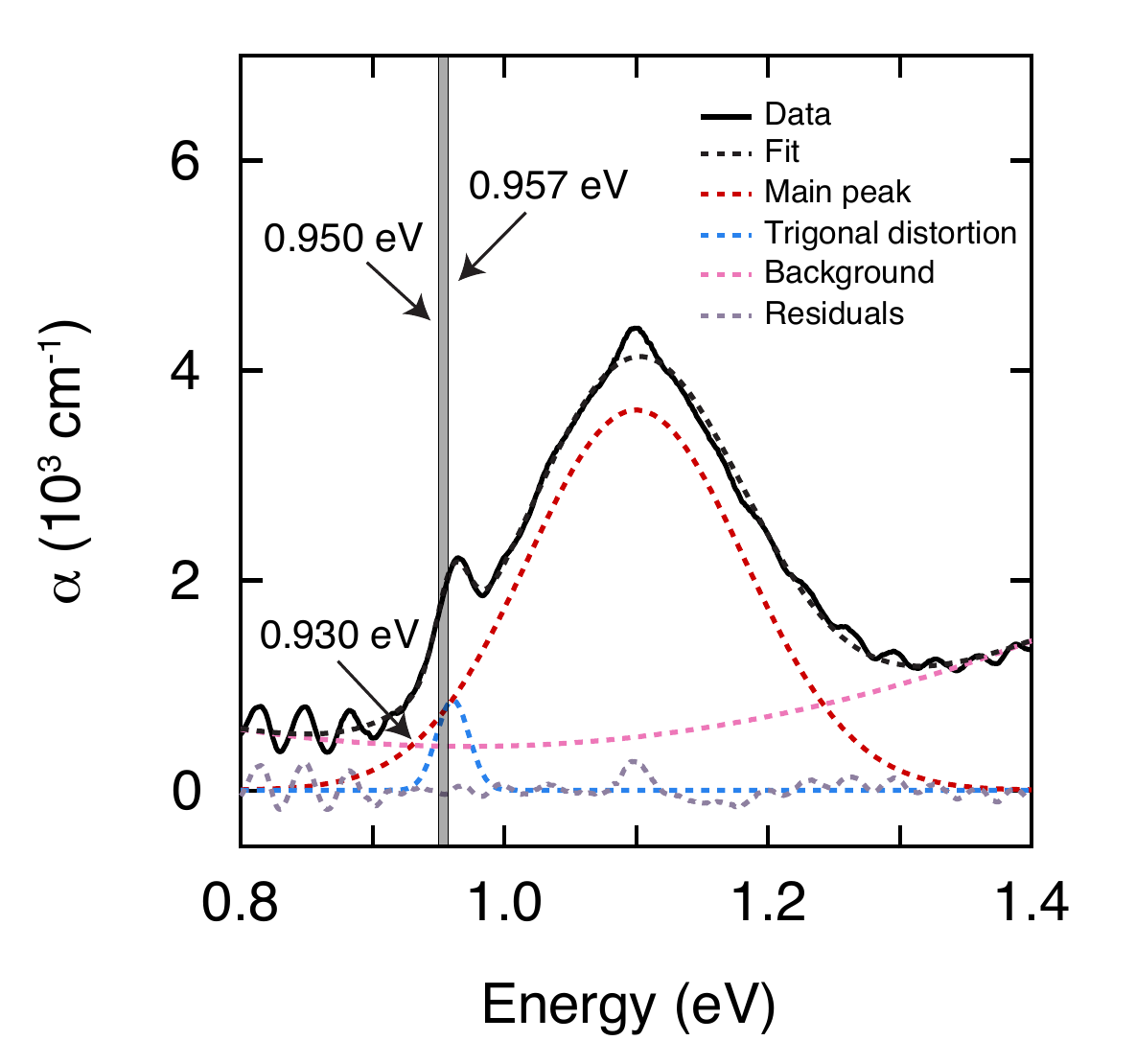}
\caption{Optical absorption ($\alpha$) of NiPS$_3$ at 20 K showing the $d$-$d$ transition around 1.1~eV. The data is fit using the following components: a Gaussian for the main $d$-$d$ transition peak, a Gaussian for the smaller lower-energy peak split due to a trigonal distortion, and a polynomial background. The vertical bar denotes the pump photon energies across which there occurs a sharp change in the pump polarization dependence of the coherent magnon amplitude. We also highlight the energy at which the main $d$-$d$ transition peak and the background have the same value of $\alpha$ (0.930~eV), which is treated as a cutoff energy between the transparent and resonant regions of photoexcitation. The optical absorption data is taken from Ref.~\cite{belvin2021exciton}.}
\label{fig:FigS1}
\end{center}
\end{figure}

\begin{figure*}
\begin{center}
\includegraphics[width=2\columnwidth]{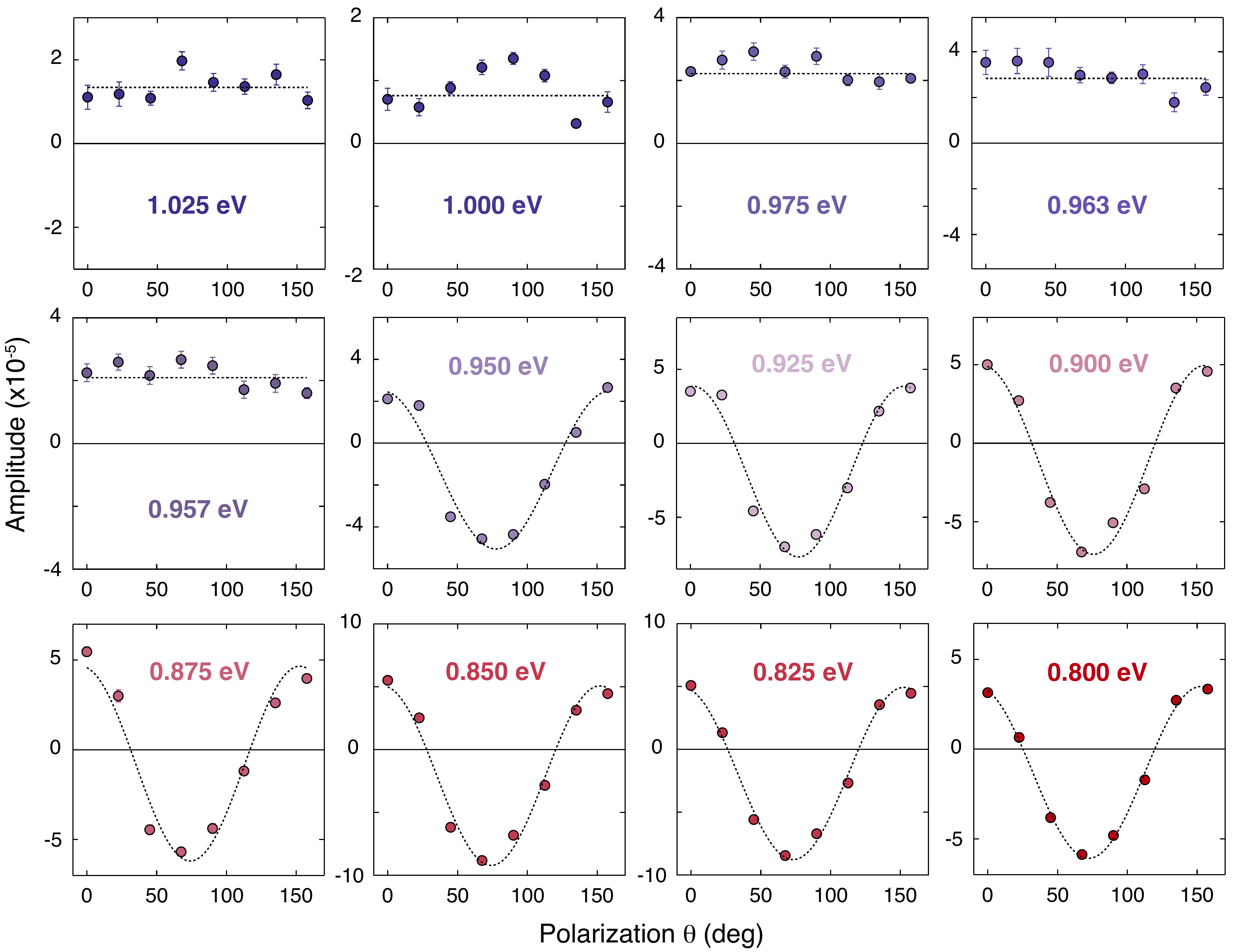}
\caption{Dependence of the amplitude of the 1.3~THz magnon mode on the pump polarization angle $\theta$ (as defined in Fig.~1 in the main text) extracted from fits to the THz emission at 15~K for a range of pump photon energies from 0.800 to 1.025~eV. An abrupt change in the generation mechanism of the coherent magnon is observed between 0.950 and 0.957~eV as the polarization dependence changes from a cos($2\theta$) behavior to (nearly) independent of polarization. The error bars represent the 95\% confidence interval from the fits (the size of the points is sometimes larger than the error bars). The absorbed pump fluence is maximized at each pump photon energy (based on the maximum power of the OPA output) to give the best signal-to-noise ratio, so the fluence is not kept constant across different photon energies.}
\label{fig:FigS2}
\end{center}
\end{figure*}

Figure~\ref{fig:FigS1} displays the optical absorption ($\alpha$) of NiPS$_3$ at 20~K in the region of the $d$-$d$ transition around 1.1~eV. The smaller peak around 0.95~eV to the left of the large peak is attributed to a splitting of the $d$-$d$ transition due to a lifting of the degeneracy of the octahedral crystal field states by a trigonal distortion \cite{piacentini1982optical,banda1986optical,grasso1986optical}. The oscillatory features present at lower energies are artifacts of the measurement setup. We fit the spectrum in Fig.~\ref{fig:FigS1} using the following components: a Gaussian function for the main $d$-$d$ transition peak, another Gaussian function for the smaller peak split due to the trigonal distortion, and a polynomial function for the background. As will be discussed in Sec.~\ref{sec:PolDeps}, the energy at which the background and the main $d$-$d$ peak have the same value of $\alpha$ (0.930~eV) is treated as a cutoff energy between the resonant and transparent regions of photoexcitation. The shaded vertical line denotes the narrow range of pump photon energies across which there occurs a sharp change in the experimentally observed pump polarization dependence of the coherent magnon amplitude (see Fig.~\ref{fig:FigS2}).

\section{Pump Polarization Dependence of the THz Emission at Additional Pump Photon Energies}
\label{sec:PolDeps}

\begin{figure*}
\begin{center}
\includegraphics[width=2\columnwidth]{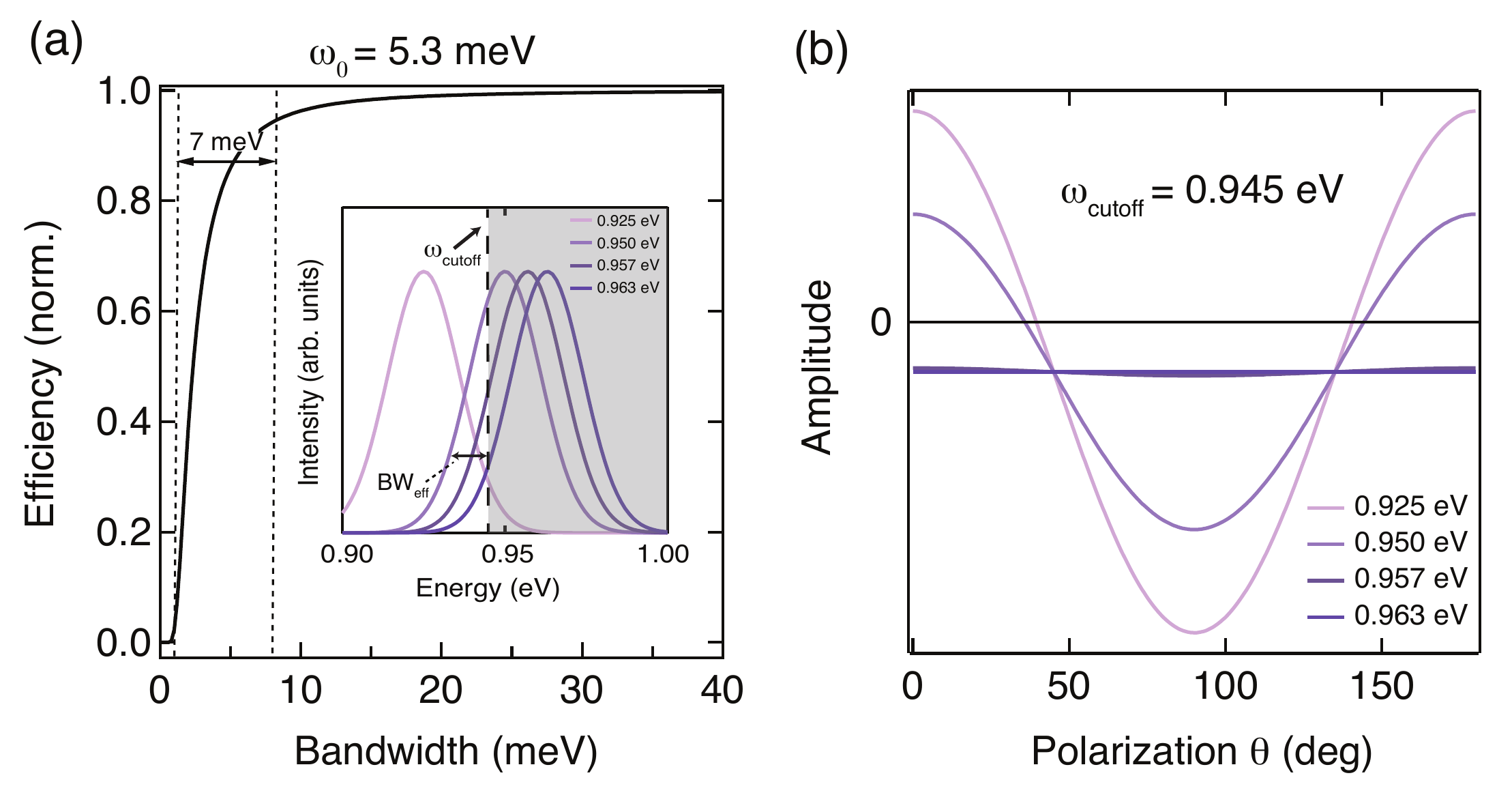}
\caption{(a) Efficiency of the off-resonant mechanism as a function of pump pulse bandwidth for the coherent generation of the 1.3~THz (5.3~meV) magnon mode. The inset illustrates the portion of the pulse bandwidth that lies below the cutoff energy and therefore contributes to this mechanism. (b) Simulation of the pump polarization dependence of the magnon amplitude using the model given by Eq.~\ref{eq:cutoff}. A sharp change from a polarization-dependent to a polarization-independent behavior is observed in agreement with the experimental results in Fig.~\ref{fig:FigS2}.}
\label{fig:FigS3}
\end{center}
\end{figure*}

In addition to the data shown in Fig.~2 in the main text, we perform a dependence of the THz emission on the linear polarization angle of the pump for the entire range of pump photon energies. We fit the raw data curves for each pump photon energy (not shown) to sinusoidal oscillations and extract the amplitude of the 1.3~THz coherent magnon mode. The results of all of the fits are shown in Fig.~\ref{fig:FigS2}. There is an abrupt change in the behavior of the magnon amplitude as a function of pump polarization angle as we cross from 0.950 to 0.957~eV. At lower photon energies when the pump is in a region of transparency, the magnon amplitude varies as cos($2\theta$), whereas at higher photon energies when the pump is resonant with the $d$-$d$ transition, the amplitude is nearly independent of the pump polarization. This markedly different behavior is indicative of distinct generation mechanisms of the coherent magnon, as discussed in the main text. The polarization dependence in transparency is a consequence of the symmetry of the electric field bilinears which follow that of the Raman tensor \cite{seifert2022ultrafast}. Explicitly, as discussed in Ref.~\onlinecite{seifert2022ultrafast}, the relevant electric field bilinears transforming in the $E$ irreducible representation of the $D_3$ symmetry group of a Ni$^{2+}$ are given by $\mathcal{E}_x \mathcal{E}_y^\ast + \mathcal{E}_x \mathcal{E}_y^\ast$ and $|\mathcal{E}_x|^2 - |\mathcal{E}_y|^2$, where $\mathcal{E}_\alpha$ are the $\alpha=x,y$-components of the light's polarization vector. Under a rotation of the polarization by an angle $\alpha$, i.e. $\mathcal{E}_x \to \mathcal{E}_x \cos \alpha + \mathcal{E}_y \sin \alpha$ and $\mathcal{E}_y \to - \mathcal{E}_x \sin \alpha + \mathcal{E}_y \cos \alpha$, these components transform as
\begin{subequations}\begin{align}
	|\mathcal{E}_x|^2 - |\mathcal{E}_y|^2 &\to  \left( \mathcal{E}_x \mathcal{E}_y^\ast + \mathcal{E}_y \mathcal{E}_x^\ast \right) \cos 2 \alpha \notag\\ &\ + \left(|\mathcal{E}_x|^2 - |\mathcal{E}_y|^2\right)\sin 2\alpha \\
	\mathcal{E}_x \mathcal{E}_y^\ast + \mathcal{E}_y \mathcal{E}_x^\ast &\to -\left(|\mathcal{E}_x|^2 - |\mathcal{E}_y|^2\right)\sin 2\alpha \notag\\ &\ + \left( \mathcal{E}_x \mathcal{E}_y^\ast + \mathcal{E}_x \mathcal{E}_y^\ast \right) \cos 2 \alpha,
\end{align}\end{subequations}
which makes evident that magneto-optical effects relying on pump-induced single-ion anisotropies (in the $E$ irreducible presentation) will exhibit a characteristic $\sim 2 \alpha$ dependence.

The abrupt change is also consistent with the theory of the dissipation-driven mechanism on resonance (see Sec.~\ref{Adding dissipation}). We present the following simple model to explain this sharp change.
In the transparent region, the magnon excitation is based on a magneto-optical modification of the single-ion anisotropy by the driving light, analogous to an impulsive stimulated Raman scattering mechanism. Guided by previous literature \cite{MerlinISRSDECP1997}, we assume the following dependence on the pump pulse bandwidth: $e^{(-(1/2)\omega_0^2\tau^2)}=e^{(-(1/8)(\omega_0/\mathrm{BW})^2)}$, where $\tau$ is the pulse duration, $\mathrm{BW}=1/(2\tau)$ is the pulse bandwidth, and $\omega_0$ is the frequency of the collective excitation.
Based on the optical absorption spectrum (see Fig.~\ref{fig:FigS1}), we posit that any pulse bandwidth above a certain photon energy will be entirely absorbed by the $d$-$d$ transition and will only contribute to the resonant dissipation-based mechanism. We expect this cutoff energy to occur around 0.930~eV, where the background and the main peak of the $d$-$d$ transition have equivalent values of the optical absorption $\alpha$. The duration of the pump pulses from our OPA is approximately 70~fs, which corresponds to a bandwidth of around 26~meV. In the vicinity of the cutoff energy, we can employ the following mathematical model for the amplitude of the coherent magnon:

\begin{widetext} 
\begin{equation}
\label{eq:cutoff}
\mathrm{Amplitude}(\theta)=\int_{0}^{\omega_\mathrm{cutoff}} I_0 e^{-1/2((\omega-\omega_c)/\mathrm{BW})^2} \,d\omega\left(e^{-1/8(\omega_0/\mathrm{BW}_\mathrm{eff})^2}\right) \Beta \cos{2 \theta} +\mathrm{Sat.},
\end{equation}
\end{widetext}

\begin{figure*}[t]
\begin{center}
\includegraphics[width=2\columnwidth]{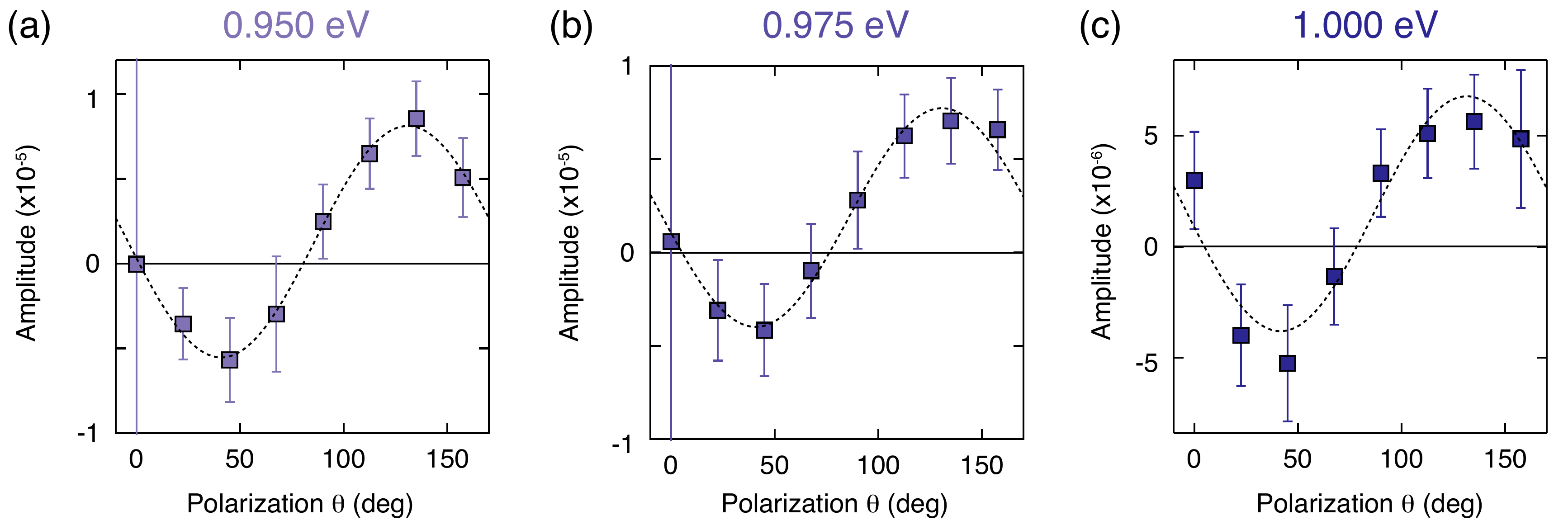}
\caption{Amplitude of the 0.3~THz mode as a function of pump polarization for pump photon energies 0.950~eV (a), 0.975~eV (b), and 1.000~eV (c) at a temperature of 15~K. The error bars represent the 95\% confidence interval from the fits. The dashed lines are fits to the form $A\,$cos$(2\theta)+C$. We note that the absorbed pump fluence is not constant as a function of pump photon energy similar to the data in Fig.~\ref{fig:FigS2}.}
\label{fig:FigS4}
\end{center}
\end{figure*}

where $\theta$ is the pump polarization angle; $I_0$, $\mathrm{BW}$, and $\omega_c$ are the initial intensity, the starting bandwidth, and the central frequency of the pump pulse, respectively; $\mathrm{BW}_\mathrm{eff}=\mathrm{BW}/2-(\omega_c-\omega_\mathrm{cutoff})$ is the effective portion of the pump pulse bandwidth that lies below the cutoff frequency $\omega_\mathrm{cutoff}$; and $\Beta$ is the Raman coupling coefficient. It is assumed that enough of the initial intensity $I_0$ of the pump pulse is present to saturate the $d$-$d$ transition, giving an offset term $\mathrm{Sat}.$ Therefore, we integrate the intensity of the pulse in the region where the off-resonant magneto-optical effect is dominant, multiply by its efficiency with respect to the effective bandwidth and by a $\cos(2\theta)$ function predicted for this off-resonant mechanism, and add the saturated $\theta$-independent term for the resonant excitation. The efficiency of the off-resonant mechanism is plotted in Fig.~\ref{fig:FigS3}(a). The inset depicts the portion of the pump pulse bandwidth that lies below the cutoff energy and thus contributes to this mechanism. We can see that the efficiency drops off rapidly when the pulse bandwidth is reduced below about 7~meV.
In Fig.~\ref{fig:FigS3}(b), we show a simulation of the polarization dependence of the magnon amplitude using the model given by Eq.~\ref{eq:cutoff}.
The exact cutoff value was adjusted to 0.945~eV to match the experimental results displayed in Fig.~\ref{fig:FigS2}. Given that there likely exist slight discrepancies due to sample variation or the exact pump photon energies and bandwidths from our commercial OPA, this simple model is a plausible explanation for the experimental observations and highlights that only a small change in the pump photon energy (much less than the bandwidth of the pump pulse) is needed to switch from a polarization-dependent to a polarization-independent magnon amplitude.

\section{Lower-Energy Magnon Mode}

A previous study on NiPS$_3$ using similar photoexcitation parameters but a different detection scheme based on Faraday rotation reported a magnon mode at a frequency of approximately 0.3~THz \cite{afanasiev2021controlling}. The authors found that this mode is only excited coherently when the pump is resonant with the $d$-$d$ transition around 1.0~eV. In our THz emission data, we do observe this 0.3~THz mode but its amplitude is only just above our detection limit. We perform fits to the THz emission traces shown in the main text using two damped sine waves to reveal this additional magnon mode. The dependence of the amplitude of the 0.3~THz mode on the pump polarization is displayed in Fig.~\ref{fig:FigS4} for pump photon energies 0.950, 0.975, and 1.000~eV. The polarization dependence exhibits a cos$(2\theta)$ behavior and closely resembles that reported in Ref.~\cite{afanasiev2021controlling}, confirming that we are detecting the same magnon mode. Additionally, similar to Ref.~\cite{afanasiev2021controlling}, we only detect this 0.3~THz mode when the pump is in resonance with the $d$-$d$ transition. However, we find that the onset of the 0.3~THz mode occurs at a slightly lower pump photon energy than the abrupt change in the polarization dependence of the 1.3~THz mode, further indicating that these two phenomena arise from independent mechanisms. We note that the 0.3~THz mode does not appear in the equilibrium THz absorption, in which only the 1.3~THz mode is present \cite{belvin2021exciton}.

Owing to the differences in the geometries of the detection methods of Faraday rotation (sensitive to out-of-plane magnetization oscillations) and THz emission (sensitive to in-plane oscillations), it is not always expected to observe the same magnon excitations with both techniques. Indeed, this can explain why the Faraday rotation study (Ref.~\cite{afanasiev2021controlling}) prominently observes the 0.3~THz mode but does not detect the 1.3~THz mode. In contrast, our THz emission data reveals the 1.3~THz mode at all pump photon energies measured (notably generated via distinct mechanisms in transparency and weak absorption, as discussed in the main text). The slight oblique angle of incidence of our pump beam ($\sim15^{\circ}$, see Sec.~\ref{sec:expdet}) likely explains how we can weakly excite the 0.3~THz mode using THz emission. Besides the 0.3~THz mode, the authors of  Ref.~\cite{afanasiev2021controlling} detect a different mode at 0.9~THz when pumping in transparency. We do not observe a 0.9~THz in THz emission mode at any photoexcitation energy. A recent Raman spectroscopy study determined that the 0.9~THz mode is in fact a zone-folded phonon rather than a magnon \cite{sun2024dimensionality}.

\begin{figure}[t]
\begin{center}
\includegraphics[width=0.9\columnwidth]{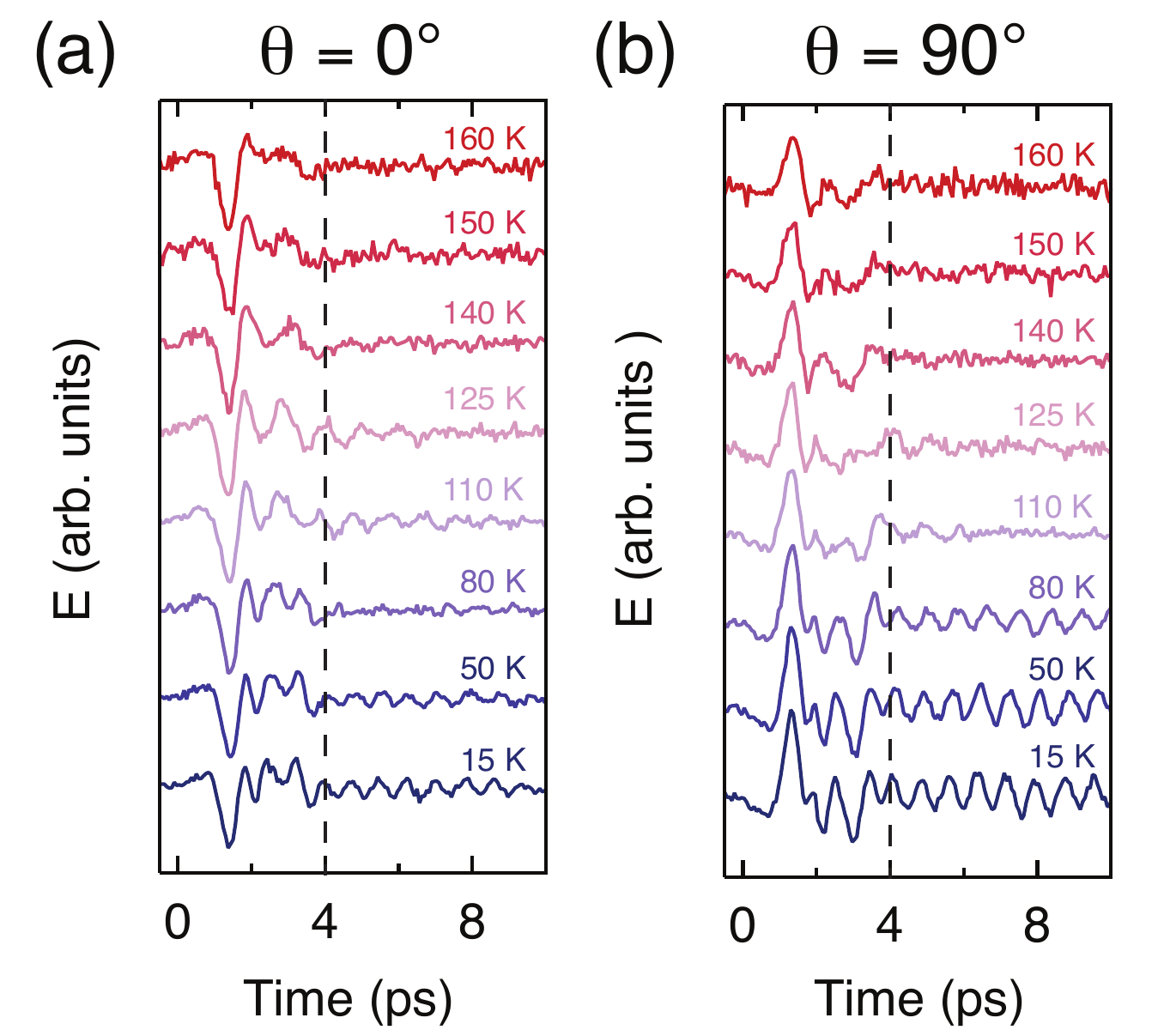}
\caption{THz emission at $\theta = 0^\circ$ (a) and $\theta = 90^\circ$ (b) at various temperatures. The initial transient signal, attributed to surface optical rectification, is present at all temperatures, including above the N\'eel temperature. Therefore, this signal is of non-magnetic origin.}
\label{fig:FigS5}
\end{center}
\end{figure}

\section{Optical Rectification Transient On Resonance}

The initial transient signal in the THz emission is attributed to surface optical rectification. This signal is present above the N\'eel temperature and it changes polarity upon a rotation of the pump polarization by 90 degrees (see Fig.~\ref{fig:FigS5}). These observations indicate that the signal is non-magnetic and it obeys the same polarization dependence observed by THz fields produced by optical rectification. This initial transient is present only when the pump is resonant with the $d$-$d$ transition (also the same pump photon energies in which the 0.3~THz magnon is observed). Since the optical rectification signal decays to zero after a few picoseconds (as seen in the higher temperature traces in Fig.~\ref{fig:FigS5} containing no magnon oscillations), we can fit the low-temperature oscillations starting after this time delay (indicated by the black dashed line) to obtain purely the magnon response.

\begin{figure}[b]
\begin{center}
\includegraphics[width=\columnwidth]{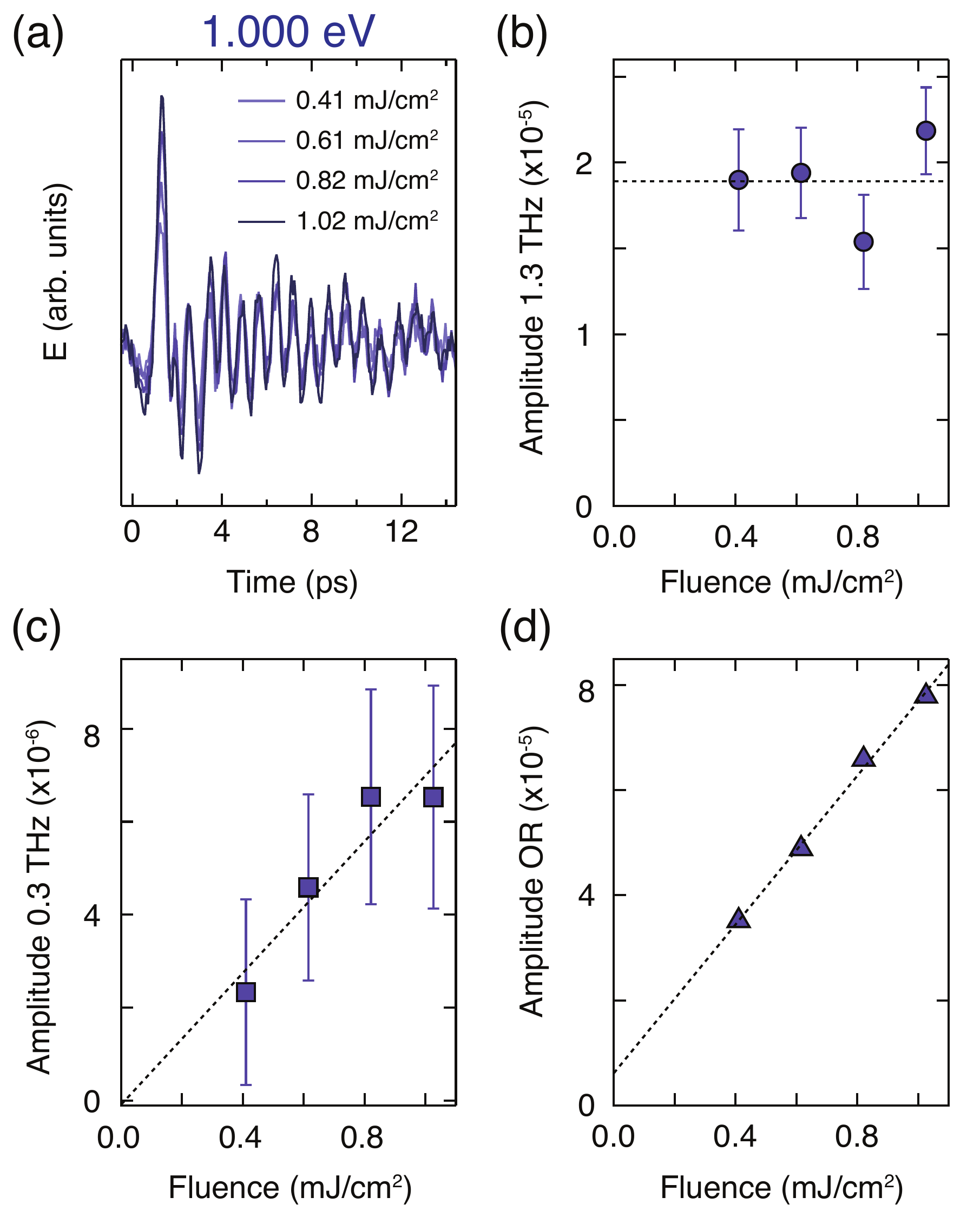}
\caption{(a) THz emission data at 15~K for various absorbed pump fluences in the high-fluence regime for a pump photon energy of 1.0 eV. Amplitudes of the 1.3~THz magnon (b), the 0.3~THz magnon (c), and the optical rectification (OR) signal (d) as a function of pump fluence. The magnon amplitudes were extracted from fits to two oscillations, and the error bars denote the 95\% confidence interval from the fits. The amplitude of the optical rectification signal is taken as the maximum value of the raw data traces.}
\label{fig:FigS6}
\end{center}
\end{figure}

\section{Pump Fluence Dependence of the THz Emission}

\begin{figure}[t]
\begin{center}
\includegraphics[width=0.92\columnwidth]{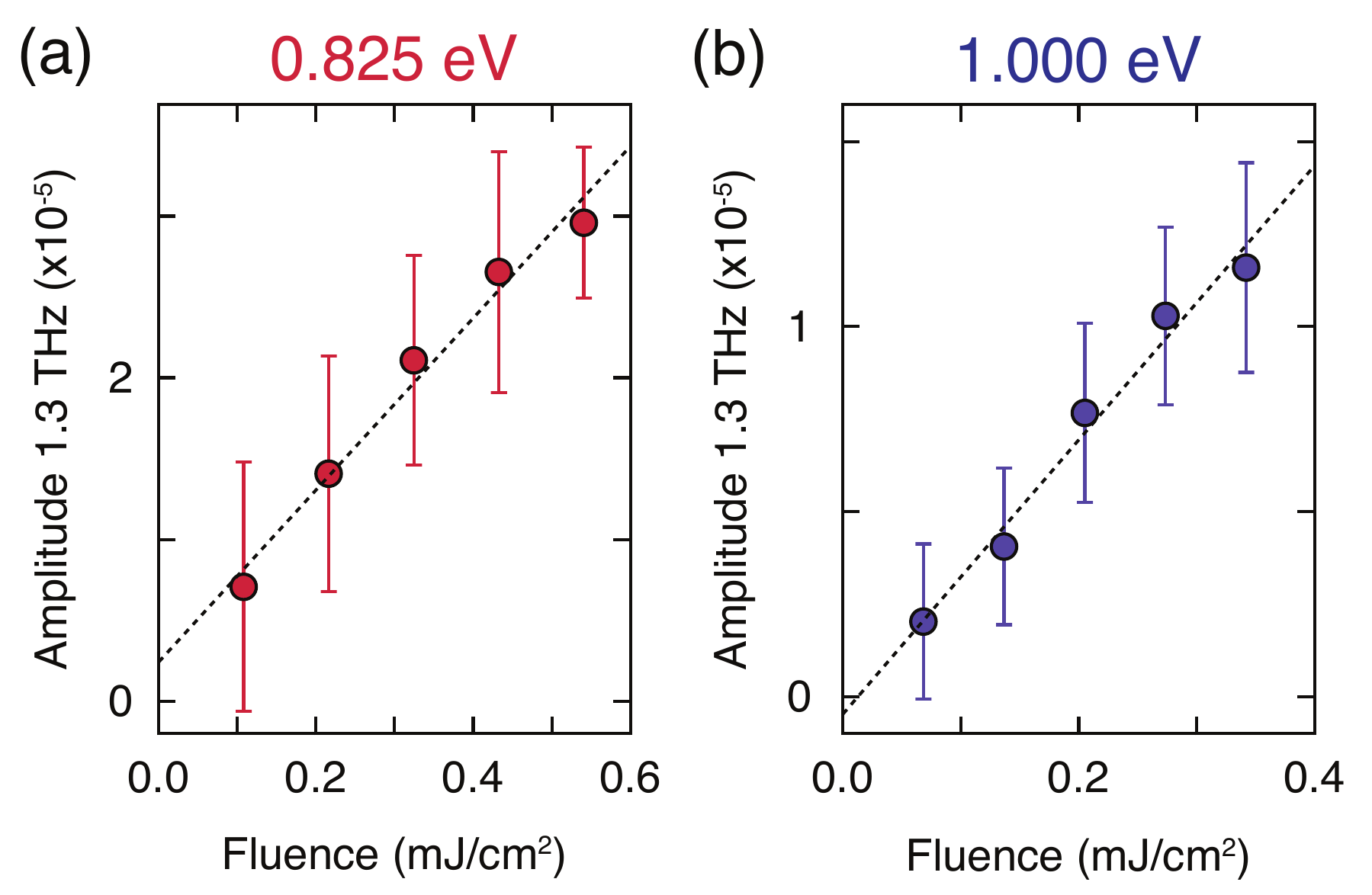}
\caption{Amplitude of the 1.3~THz magnon mode as a function of absorbed pump fluence at 15~K in the low-fluence regime for pump photon energies 0.825~eV (a) and 1.000~eV (b) at a fixed pump polarization of $\theta = 67.5^\circ$ and $\theta = 0^\circ$, respectively.}
\label{fig:FigS7}
\end{center}
\end{figure}

As mentioned in the main text, to further understand the different generation mechanisms of the coherent magnon in transparency and resonant with the $d$-$d$ transition, we examine the magnon amplitude as a function of absorbed pump fluence. First, we consider resonant photoexcitation at a pump photon energy of 1.0~eV (Fig.~\ref{fig:FigS6}). We choose the pump polarization to be $\theta = 112.5^\circ$, where the amplitude of the 0.3~THz magnon mode is nearly maximized, so that we can capture the behavior of both modes. Figure~\ref{fig:FigS6}(a) displays the raw data of the THz emission at several fluences large enough to achieve a sufficient signal-to-noise ratio to observe the 0.3~THz mode (we denote this the high-fluence regime). The amplitudes of the 1.3~THz and 0.3~THz modes extracted from the fits are shown in Figs.~\ref{fig:FigS6}(b) and (c), respectively, and the amplitude of the optical rectification signal (taken as the maximum value of each trace) is displayed in Fig.~\ref{fig:FigS6}(d). With increasing fluence, the amplitudes of the 0.3~THz mode and the optical rectification signal rise linearly, whereas the amplitude of the 1.3~THz mode appears to be already saturated.

We also examine the behavior of the 1.3~THz magnon in the low-fluence regime for photoexcitation in both transparency and resonant with the $d$-$d$ transition. In both cases, we observe that the amplitude grows linearly with increasing pump fluence (see Fig.~\ref{fig:FigS7}). For the measurement on resonance, we use a pump polarization of $\theta = 0^\circ$ so that the 0.3~THz mode is minimized and we can more accurately extract solely the contribution of the 1.3~THz mode.

Combining the response of the 1.3~THz magnon in the low- and high-fluence regimes on resonance, we see that the amplitude first rises linearly and then saturates, which is consistent with our theoretical model that takes into account the effects of dissipation (see the discussion in the main text and Sec.~\ref{sec:TheoryResults}). We note that the measurements in the low- and high-fluence regimes were performed on different days and therefore cannot be directly compared in a quantitative manner. In transparency, the linear fluence dependence of the 1.3~THz magnon is as expected for an impulsive stimulated Raman scattering generation mechanism.

\begin{figure}[b]
\begin{center}
\includegraphics[width=0.95\columnwidth]{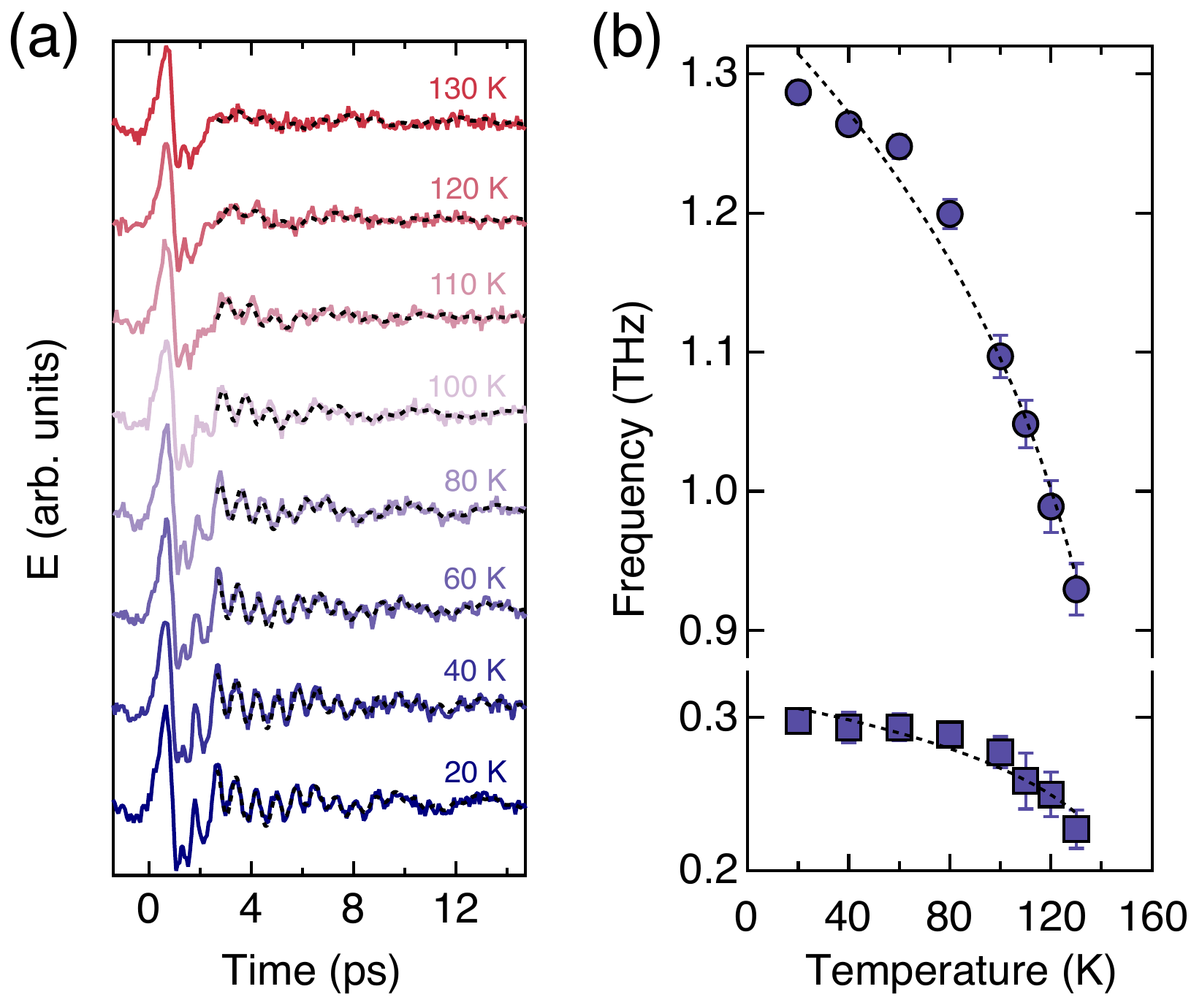}
\caption{(a) THz emission as a function of temperature using a pump photon energy of 1.0~eV. The pump polarization is fixed at $\theta=45^\circ$ and the absorbed pump fluence is 0.74~mJ/cm$^2$. The raw data are fit with two damped sinusoids (dashed black lines). (b) Frequencies of the two modes as a function of temperature extracted from the fits. The error bars represent the 95\% confidence interval of the corresponding fit parameters. The frequency versus temperature plots are then fit to an order parameter curve of the form $(T_N-T)^\beta$, where $T_N$ is fixed at 157~K (see Sec.~\ref{sec:expdet} for the determination of $T_N$) and $\beta$ is determined to be $0.21 \pm 0.01$ and $0.16 \pm 0.02$ for the 1.3~THz and 0.3~THz modes, respectively.}
\label{fig:FigS8}
\end{center}
\end{figure}

\section{Temperature Dependence of the THz Emission}

To verify that the observed coherent collective mode at 1.3~THz is a magnon excitation, we perform a temperature dependence of the THz emission signal at a pump photon energy of 1.0~eV. Figure~\ref{fig:FigS8}(a) shows the raw data of the THz emission as a function of temperature. We fit each trace to two damped sine waves (to also account for the presence of the lower-energy 0.3~THz mode), and the extracted frequencies at each temperature are displayed in Fig.~\ref{fig:FigS8}(b). The frequency of the 1.3~THz mode (circular points) softens with a clear order parameter behavior towards $T_N$ and the mode is absent above this temperature. Further evidence of the magnetic origin of this mode is given by its extremely sharp linewidth in the equilibrium THz absorption and its small oscillator strength indicating a magnetic dipole allowed transition (see Ref.~\cite{belvin2021exciton}).

The frequency of the 0.3~THz mode (square points in Fig.~\ref{fig:FigS8}(b)) also displays a softening towards $T_N$ with an order parameter trend and the mode is no longer present above the transition temperature, suggesting that it is also a magnetic excitation, in agreement with Ref.~\cite{afanasiev2021controlling}. A recent inelastic neutron scattering study on NiPS$_3$ also confirmed that the two low-energy zone-center magnons in this material are the 0.3 and 1.3~THz modes \cite{scheie2023spin}.

\section{THz Transmission}

\begin{figure}[t]
\begin{center}
\includegraphics[width=0.95\columnwidth]{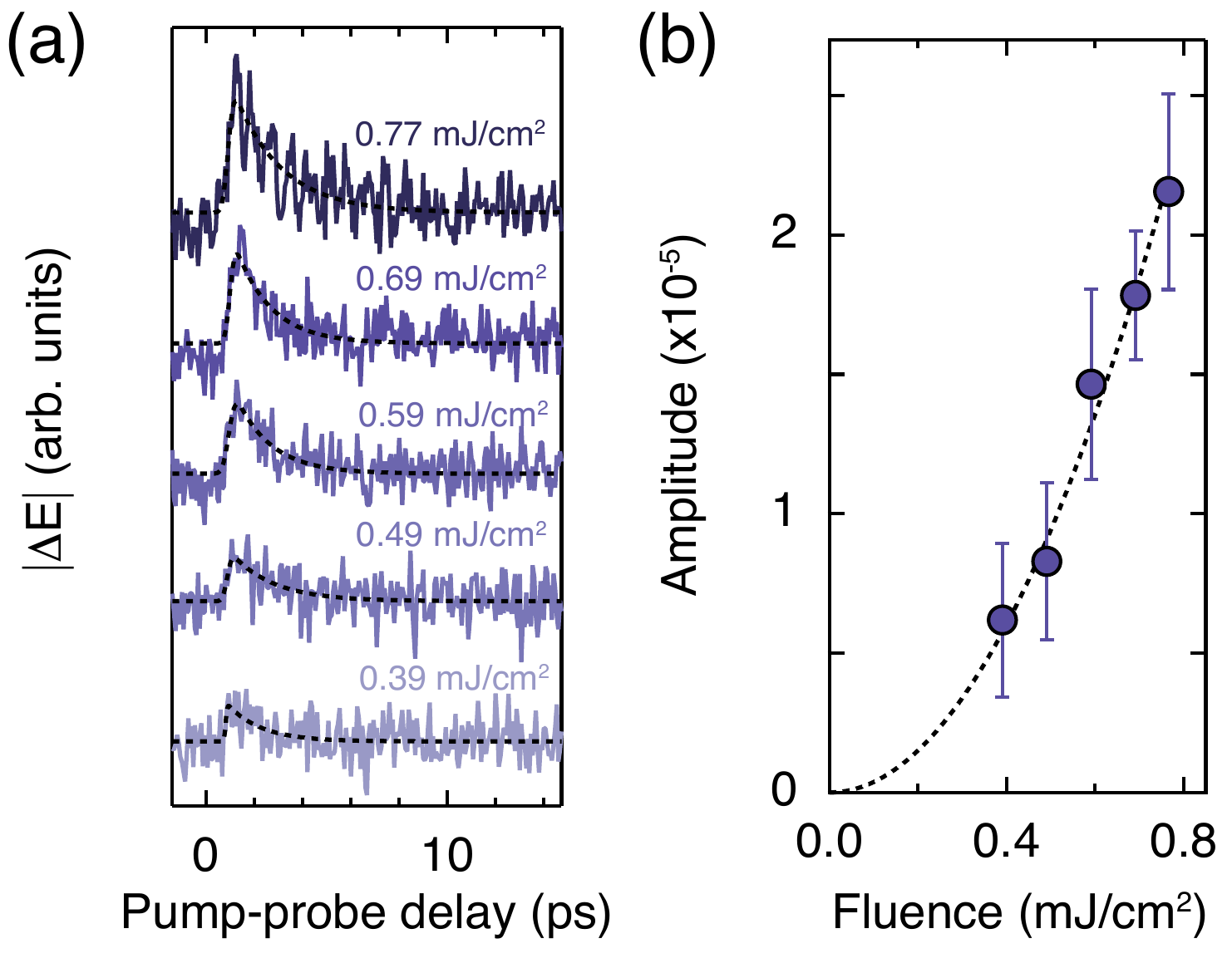}
\caption{(a) Absolute value of the pump-induced change in the transmitted THz electric field ($\Delta E$) after subtracting the THz emission signal for various absorbed pump fluences. The pump photon energy is 1.0~eV and the temperature is 20~K. Each curve is fit to a decaying exponential (black dashed lines). (b) Amplitude of the exponential extracted from the fits in (a). The amplitude varies quadratically with absorbed pump fluence. The error bars denote the 95\% confidence interval from the fits.}
\label{fig:FigS9}
\end{center}
\end{figure}

In addition to the THz emission data presented in the main text and above sections, we also perform THz transmission measurements. The details of this technique are described in Ref.~\cite{belvin2021exciton}. Whereas THz emission spectroscopy can detect the oscillating dipole radiation from coherent collective modes, THz transmission measurements are useful for observing the presence of charge carriers. Figure~\ref{fig:FigS9}(a) shows the pump-induced change in the THz electric field ($\Delta E$) as a function of pump-probe delay after subtracting the THz emission traces at the same fluences (to eliminate the coherent magnon contribution) for a pump photon energy of 1.0~eV. The signal displays an increase followed by a decaying exponential, a typical signature of the production of charge carriers, which absorb THz light and undergo subsequent relaxation dynamics. A nonzero difference between the THz transmission and emission signals is only observed when the pump is resonant with the $d$-$d$ transition, indicating that charge carriers are produced only under resonant photoexcitation.

To determine the dependence of the charge carriers on the absorbed pump fluence, we fit each trace in Fig.~\ref{fig:FigS9}(a) to an exponential function and extract its amplitude, shown in Fig.~\ref{fig:FigS9}(b). The amplitude is quadratic in fluence, suggesting that the carriers are produced via exciton dissociation (microscopic mechanisms include exciton-exciton collision ionization \cite{braun1968singlet,lee1993transient} and pump-induced exciton photoionization \cite{bergman1974photoconductivity}), similar to what is observed in Ref.~\cite{belvin2021exciton}. This can be understood considering that the $d$-$d$ transition around 1.1~eV is a Frenkel exciton that is localized on the Ni$^{2+}$ ion and has a strong binding energy. We therefore expect only a small fraction of excitons to dissociate, consistent with the small amplitude of the exponential signal (an order of magnitude smaller than that of Ref.~\cite{belvin2021exciton} in which the excitons are less tightly bound). Since no exponential signal is observed for pump photon energies in the transparent region, it is unlikely that the carriers are generated through direct multi-photon absorption via a virtual state.

We remark that the component of the THz transmission signal due to the charge carriers is very small and therefore a frequency-resolved measurement does not yield a sufficient signal-to-noise ratio to perform a reliable analysis of the nature of the carriers.

\section{Theory: General setup}

We consider an isolated Ni$^{2+}$ ion in an octahedral crystal field that is driven in resonance with the $d$-$d$ transition from the $A_{2g}$ to $T_{2g}$ multiplets. When driven on resonance, the system can be modeled within a Lindbladian picture, where the system undergoes \emph{both} unitary time evolution (induced by driving with the electric field of the pump pulse) as well as dissipative dynamics due to the spontaneous decay from the excited state back to the ground-state manifold.

\subsection{Unitary time evolution}\label{Theory:Setup}

\subsubsection{Light-matter interaction}

For NiPS$_3$, the ground state is spanned by $\ket{A_{2g},S^z=\pm 1,0}$ and the excited states by $\ket{J_\mathrm{eff},J_\mathrm{eff}^z}$ with $J_\mathrm{eff} = 0,1,2$.
In the following, we denote the ground-state levels by index $s$ and the excited states by index $m$. The Hamiltonian of the single ion interacting with the electric field of the driving light is then given by
\begin{equation}
	H = H_0 + V(t)
\end{equation}
where the undriven system is governed by the Hamiltonian 
\begin{equation}
	H_0 = E_0 \sum_s \ket{s} \bra{s} + \sum_m E_m \ket{m} \bra{m},
\end{equation}
written in its diagonal eigenbasis. We use $\ket{s}$ as a short form for the three states $\ket{A_{2g},S^z=\pm 1,0}$ spanning the threefold degenerate ground-state manifold (with some ground-state energy $E_0$), and $\ket{m}$ denote various excited states with energies $E_m$. $V$ denotes the light-matter interaction $V(t) = V(\omega) \eu^{\iu \omega t} + V(\omega)^\ast \eu^{-\iu \omega t}  \equiv r_\alpha \mathcal{E}_\alpha \eu^{\iu \omega t} + r_\alpha \mathcal{E}_\alpha^\ast \eu^{-\iu \omega t}$. Here, $r^\alpha$ with $\alpha=x,y$ are components of the position operator (matrix elements of which yield transition dipole moments), and $\mathcal{E}_\alpha(\omega)$ are components of the light's (complex) polarization vector at frequency $\omega$.
The system is described by some density matrix $\rho$. We work in the interaction picture $\rho^I = \eu^{\iu H_0 t} \rho \eu^{-\iu H_0 t}$ where the time evolution is determined by
\begin{equation}
	\iu \partial_t\rho^I = [ V_I,\rho^I ],
\end{equation} 
and $V_I$ can be written in the form
\begin{widetext}
\begin{align}
	V_I &= \sum_{s,m} \left(\ket{s}\bra{m} \eu^{\iu (E_0 - E_m)t} \braket{s|V(t)|m} + \ket{m}\bra{s} \eu^{-\iu (E_0 - E_m)t} \braket{m|V(t)|s} \right) \\
	&\approx \sum_{s,m} \left( \ket{s}\bra{m} \eu^{\iu (\omega- \Delta_m)t} V_{sm} + \ket{m} \bra{s} \eu^{-\iu (\omega-\Delta_m)t} V^\ast_{ms} \right),
\end{align}
\end{widetext}
where $\Delta_m = E_m - E_0$, $V_{sm} = \braket{s|V(\omega)|m}$, and we have made the rotating wave approximation (i.e. dropped terms which oscillate at frequencies $\omega+\Delta_m$).
We can then write down the equations of motion for the respective sectors of the density matrix:
\begin{widetext}
\begin{align} \label{eq:only-unitary-RWA}
	\partial_t \rho^I_{s,s'} &= -\iu \sum_m \left(V_{sm} \rho^I_{ms'} \eu^{\iu (\omega-\Delta_m) t} - \rho^I_{sm} V^\ast_{ms'} \eu^{-\iu (\omega-\Delta_m) t} \right) \\
	\partial_t \rho^I_{m,s} &= -\iu \sum_{s'} V^\ast_{m,s'} \rho^I_{s',s} \eu^{-\iu (\omega-\Delta_m)t} + \iu \sum_{m'} \rho^I_{m,m'} V^\ast_{m',s} \eu^{-\iu (\omega-\Delta_m')t} \\
	\partial_t \rho^I_{s,m} &= -\iu \sum_{m'} V_{s,m'} \rho^I_{m',m} \eu^{\iu (\omega-\Delta_{m'})t} + \iu \sum_{s'} \rho^I_{s,s'} V_{s',m} \eu^{\iu (\omega-\Delta_m) t}  \\ 
	\partial_t \rho_{m,m'}^I &=  -\iu \sum_s \left(V^\ast_{ms} \rho^I_{sm'} \eu^{-\iu (\omega-\Delta_m)t} - \rho^I_{m,s} V_{s,m'} \eu^{\iu (\omega-\Delta_{m'})t } \right).
\end{align}
\end{widetext}

Note that at this stage, the $\ket{s=\pm1,0}$ states are completely degenerate. In NiPS$_3$, exchange interactions and a weak in-plane easy-axis anisotropy stabilize a zigzag-ordered ground state with the local moments pointing parallel/antiparallel to the $\hat{x}$ axis.
We capture this by a mean exchange field acting on the local moment that explicitly stabilizes the $\ket{\pm} = (\ket{1}\pm\sqrt{2}\ket{0} + \ket{-1})/2$, i.e. we add the perturbation 
\begin{equation} \label{eq:v-ex}
	V_{\pm x}^I = - (\pm h_\mathrm{ex}) S^x
\end{equation}
to the Hamiltonian. Note that this is already written in the interaction picture.
This implies that in the equation of motion for $\rho$ there is an additional term
\begin{equation}
	\partial_t \rho_{s,s'} = \dots - \iu \sum_{s''} \left[ (V^{\pm x})_{s,s''} \rho_{s'',s'} - \rho_{s,s''} (V^{\pm x})_{s'',s'}   \right].
\end{equation}

\subsection{Adding dissipation}\label{Adding dissipation}

\begin{figure*}[t]
\begin{center}
\includegraphics[width=1.9\columnwidth]{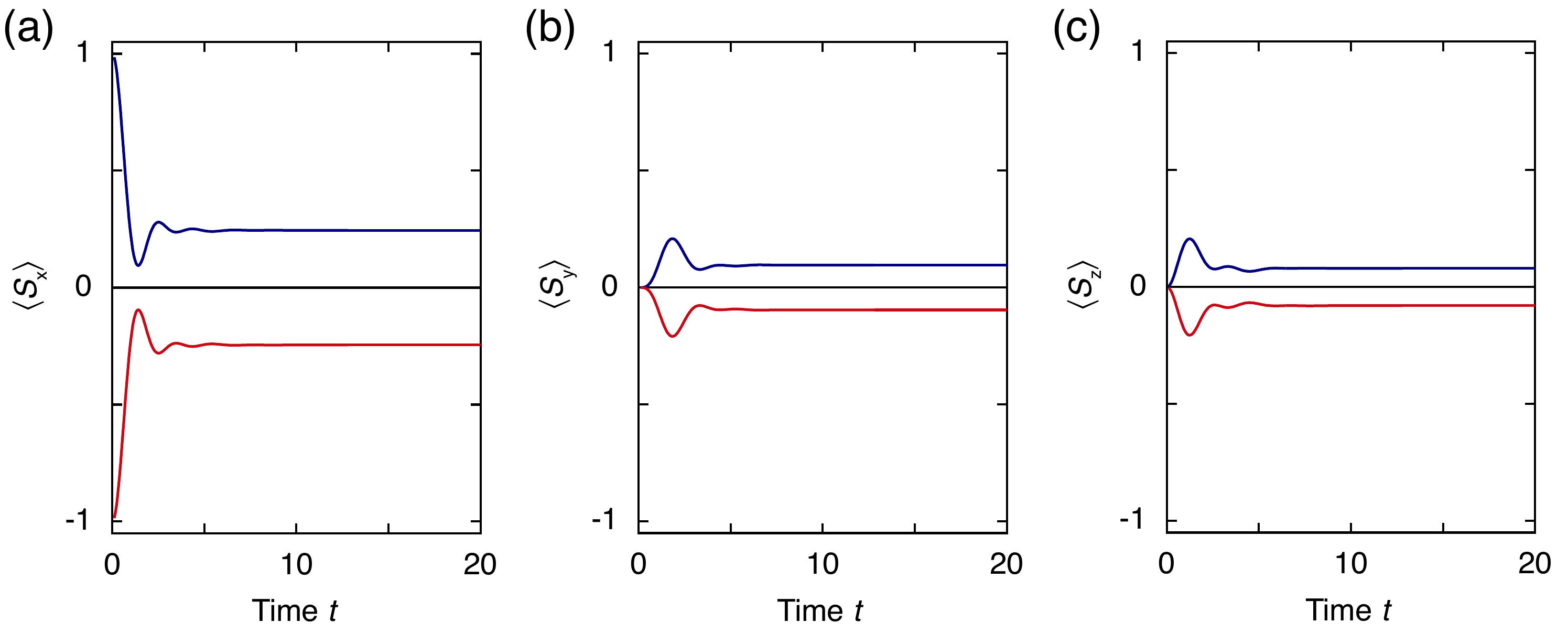}
\caption{Time evolution, starting with initial states $\ket{+}$ (blue) and $\ket{-}$ (red), with electric field amplitudes $\mathcal{E}_x = \cos \pi/6$ and $\mathcal{E}_y = \sin \pi/6$ and $\Gamma_2 =1$ (we work in units of $\Gamma_1 =1$).}
\label{fig:FigS10}
\end{center}
\end{figure*}

We now account for non-unitary time evolution. Assuming that the system's environment (bath) is Markovian, the equation of motion for the density matrix is given by the Lindblad equation
\begin{equation} \label{eq:qma}
	\partial_t \rho = \iu [\rho,H] + \sum_j \gamma_j \left[L_j \rho L_j^\dagger - \frac{1}{2} \{L_j^\dagger L_j,\rho \} \right],
\end{equation}
where the $L_j$ are jump operators for spontaneous transitions, and the $\gamma_j \geq 0$ are decay rates which must be positive by construction.
For spontaneous decay from the excited manifold to the ground state, the $L_j$ are zero except for $\braket{s|L_j|m}$.
We can transform Eq.~\eqref{eq:qma} into the interaction picture and obtain
\begin{equation} \label{eq:rho-i-l}
	\partial_t \rho^I = \iu[\rho^I,V_I] + \sum_j \gamma_j \left[ L^I_j \rho^I {L^I_j}^\dagger - \frac{1}{2} \{{L^I_j}^\dagger L^I_j,\rho \} \right],
\end{equation}
where $L_j^I = \eu^{\iu H_0 t} L_j \eu^{-\iu H_0t}$.

\subsubsection{Decay of excited states ($E$ manifold)}

\begin{figure}[t]
\begin{center}
\includegraphics[width=0.95\columnwidth]{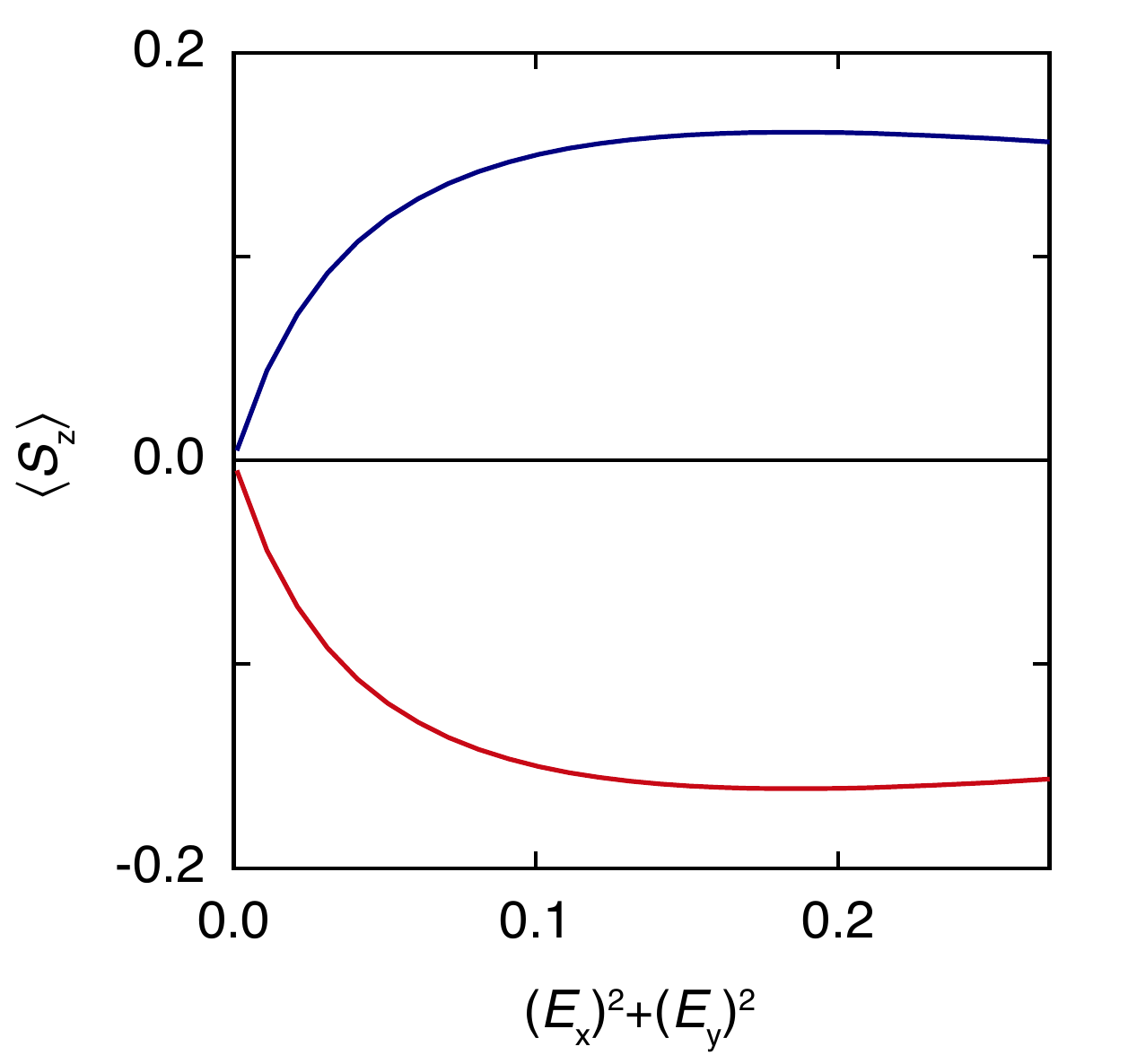}
\caption{Staggered $S^z$ at long times (steady state, $t = 1000$) as a function of the electric field amplitude with $\mathcal{E}_x / \mathcal{E}_y = \cot \pi/6$ and $\Gamma_2 =1$ (we work in units of $\Gamma_1 =1$).}
\label{fig:FigS11}
\end{center}
\end{figure}

Now we focus on the states with $J_\mathrm{eff} = 2$, which are within the $^3E_g$ manifold. This is too large of a degeneracy if we only have $C_3 \times C_2'$ symmetry, which admits at most a three-dimensional irreducible representation. We therefore impose cubic symmetry on the $J_\mathrm{eff}=2$ states. Then, these states are further split into $T_1$ and $E$ representations, which can be labeled by the $C_3$ (along the trigonal axis $\bar{z}$) eigenstates $\ket{1},\ket{\omega}$,$\ket{\omega^2}$ and $\ket{\omega},\ket{\omega^2}$, respectively. We first assume that the jump operators are symmetric with respect to the symmetry group $G=C_3 \times C_2'$ of a single ion, i.e. they satisfy $L=U_G^\dagger L U_G$, where the $U_G$ are the representation matrices of $G$ acting on the $T_1$, $E$ states. Then, the symmetry-constrained jump operator reads $L= z \left( \ket{s=+1}\bra{\omega^2}+\ket{s=-1}\bra{\omega} \right)$ with the matrix representation
\begin{equation}
	L = \begin{pmatrix}
		0 & 1 \\ 0 & 0 \\ 1 & 0 
	\end{pmatrix}
\end{equation}
such that (in the interaction picture, dropping the respective label)

\begin{equation}
	L \rho L^\dagger = |z|^2 \begin{pmatrix}
		\rho_{5,5} & 0 & \rho_{5,4} & 0 & 0 \\
		0 & 0 & 0 & 0 & 0 \\
		\rho_{4,5} & 0 & \rho_{4,4} & 0 & 0 \\
		0 & 0 & 0 & 0 & 0 \\ 
		0 & 0 & 0 & 0 & 0
	\end{pmatrix}
\end{equation}

and

\begin{equation}
	\{{L^I_j}^\dagger L^I_j,\rho \} = |z|^2 \begin{pmatrix}
0 & 0 & 0   & \rho_{1,4} & \rho_{1,5} \\
0 & 0 & 0   & \rho_{2,4} & \rho_{2,5} \\
0 & 0 & 0   & \rho_{3,4} & \rho_{3,5} \\
0 & 0 & 0   & \rho_{4,4} & \rho_{4,5} \\
\rho_{4,1} & \rho_{4,2} & \rho_{4,3} & 2 \rho_{4,4} & 2\rho_{4,5} \\
\rho_{5,1} & \rho_{5,2} & \rho_{5,3} & 2\rho_{5,4} & 2\rho_{5,5}		
	\end{pmatrix}
\end{equation}
and hence the dissipative part has the form
\begin{widetext}
\begin{equation}
	\partial_t \rho = \dots + \Gamma_1 \begin{pmatrix}
		\rho_{5,5} & 0 & \rho_{5,4} & - \rho_{1,4}/2 & - \rho_{1,5}/2 \\
		0 & 0 & 0 & -\rho_{2,4}/2 & -\rho_{2,5}/2 \\
		\rho_{4,5} & 0 & \rho_{4,4} & - \rho_{3,4}/2 & - \rho_{3,5}/2  \\
		- \rho_{4,1} /2 & -\rho_{4,2}/2 & -\rho_{4,3}/2 & -\rho_{4,4} & -\rho_{4,5} \\
		- \rho_{5,1} /2 & -\rho_{5,2}/2 & -\rho_{5,3}/2 & -\rho_{5,4} & -\rho_{5,5}
	\end{pmatrix}, 
\end{equation}
\end{widetext}
where we write $\Gamma_1 = \gamma_1 |z|^2$.

\subsubsection{Relaxation within the $S=1$ ground-state manifold}

In equilibrium (without driving), the Hamiltonian $H=H_0 + V_{\pm x}^I$ conserves  $S^x$, implying that a spin antialigned with its exchange field is a stable steady state.
To allow for relaxation back into the ground state (alignment along the exchange field), we add a second jump operator $L_2^\pm = S^y \pm \iu S^z $ which act as raising/lowering operators for $S^x$ with $[S^x, S^y \pm \iu S^z] = \pm (S^y \pm \iu S^z)$. The $+$ or $-$ sign is picked depending on the orientation of the local exchange field Eq.~\eqref{eq:v-ex}.
The corresponding relaxation rate will be called $\Gamma_2$.

\section{Theory: Computational details on time evolution} \label{sec:time-evol-naive}

\begin{figure}[b]
\begin{center}
\includegraphics[width=0.95\columnwidth]{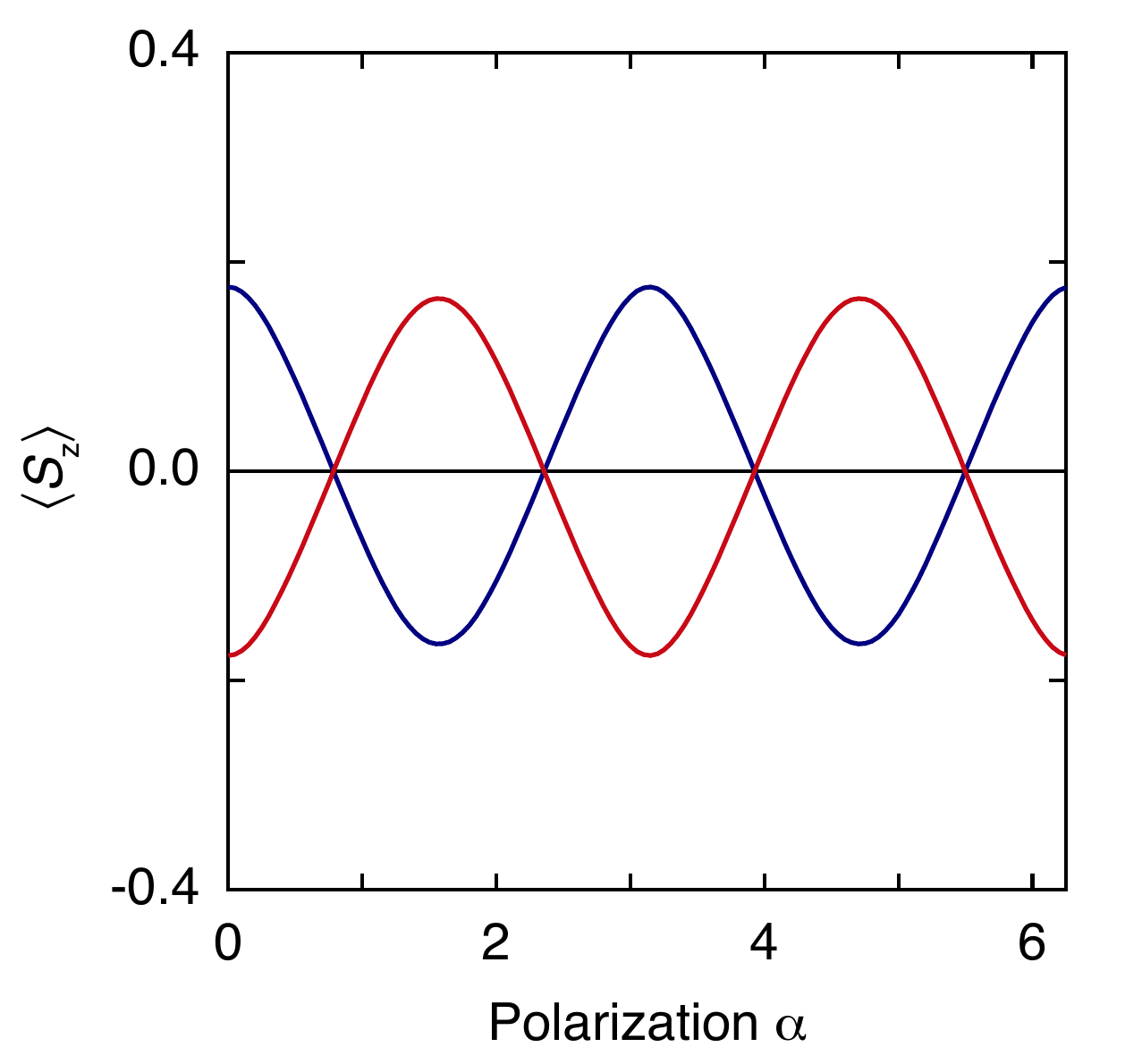}
\caption{Staggered $S^z$ with $L=((0,1),(0,0),(1,0))$, with $C_3 \times C_2$-symmetry allowed dissipation, at long times (steady state, $t = 1000$) as a function of the polarization angle $\alpha$ in $\mathcal{E}_x = \cos \alpha$ and $\mathcal{E}_y = \sin \alpha$, and $\Gamma_2 = 1$ (working in units of $\Gamma_1 = 1$).}
\label{fig:FigS12}
\end{center}
\end{figure}

We use \texttt{Mathematica} to rewrite Eq.~\eqref{eq:rho-i-l} in a vectorial form where we use the ordering $\uvec{\rho}=(\rho_{1,1},\rho_{1,2},\dots,\rho_{5,4},\rho_{5,5})^\top$.
Then, Eq.~\eqref{eq:rho-i-l} can be written of the form
\begin{equation}
	\partial_t \uvec{\rho} = \umat{W} \uvec{\rho}
\end{equation}
and the general solution is written of the form $\uvec{\rho}(t) = \eu^{\umat{W} t} \uvec{\rho}(0)$ with some initial condition $\uvec{\rho}(0)$. The matrix $\eu^{\umat{W} t}$ may be found by working in the eigenspace of $\umat{W}$ which has eigenvalues $\lambda_i$ such that
\begin{equation}
	\eu^{\umat{W} t} = \umat{U}\begin{pmatrix}
		\eu^{\lambda_1 t} & & \\
		& \ddots & \\
		&  & \eu^{\lambda_25 t} 
	\end{pmatrix} \umat{U}^{-1},
\end{equation}
where $\umat{U}$ is a matrix that contains the right eigenvectors of $\umat{W}$ as columns. Note that $\umat{W}$ is not necessarily Hermitian, so $\umat{U}$ need not be unitary, and in general we have $\lambda_i \in \mathbb{C}$.
However, we stress that $\Re[\lambda_i] \leq 0$, as guaranteed by the positivity of the master equation.
Note that for each eigenvalue $\lambda_i$, there are also left eigenvectors. These are the rows of $\umat{U}^{-1}$, or columns of $(\umat{U}^{-1})^\top$.

Focusing on time evolution by coupling the ${}^3 A_{2g}$ ground state to the $E$ states, we find that most $\lambda$ are complex.
We find that in general with $E^x,E^y \neq 0$ and $\Gamma_1,\Gamma_2 \neq 0$, there is exactly one eigenvalue with $\lambda = 0$, which corresponds to a steady state solution.

We initialize the time evolution with the pure states $\rho^\pm = \ket{S^x =\pm1} \bra{S^x=\pm 1}$ corresponding to the two inequivalent sites.

\section{Theory: Results} \label{sec:TheoryResults}

We find that \emph{without $\Gamma_2$}, i.e. without relaxation \emph{within} the $S=1$ ground-state manifold, the steady state solution $\rho_\mathrm{SS} =\ket{\psi_\mathrm{SS}}\bra{\psi_\mathrm{SS}}$ is a pure state which has vanishing spin expectation values
	$\braket{\psi_\mathrm{SS} |S^x| \psi_\mathrm{SS}} = \braket{\psi_\mathrm{SS} |S^y| \psi_\mathrm{SS}} = \braket{\psi_\mathrm{SS} |S^z| \psi_\mathrm{SS}} \equiv 0$,
implying that the local moment is in a purely quadrupolar state. An intuitive picture is that the pump primarily induces single-ion anisotropies of the form $\sim (S^\alpha)^2$, as found for the off-resonant case in Ref.~\cite{seifert2022ultrafast}.
A static Hamiltonian containing these terms may have a purely quadrupolar ground state.

Now, for a finite $\Gamma_2 \neq 0$ decay rate, we find that in the steady state, sites with ``$\pm$''-exchange fields maintain a finite magnetization along $\pm \hat{x}$, and simultaneously develop staggered $S^y$ and $S^z$ components (see Fig.~\ref{fig:FigS10}).
Therefore, the staggered $S^z$ component is crucial for exciting the 1.3~THz magnon mode.

We also plot $\langle S^z \rangle(t)$ for large times $t$ as a function of the electric field strength $(\mathcal{E}_x)^2 + (\mathcal{E}_y)^2$ in Fig.~\ref{fig:FigS11}. There is an initially ``fast'' onset and a maximum is reached for some field strength $| \mathcal{E}_\mathrm{crit} |$, after which $\langle S^z \rangle$ (weakly) decreases with increasing $|\mathcal{E}|$.
For sufficiently strong fields $|\mathcal{E}| \to \infty$, we eventually find $\langle S^\alpha \rangle = 0 \ \forall \alpha$, corresponding to the quadrupolar state that is found for $\Gamma_2 = 0$.
It can be seen that the position of $| \mathcal{E}_\mathrm{crit} |$ is controlled by $\Gamma_2$, with
$| \mathcal{E}_\mathrm{crit} |/\Gamma_1 \sim \Gamma_2/\Gamma_1$.

\begin{figure}
\begin{center}
\includegraphics[width=0.95\columnwidth]{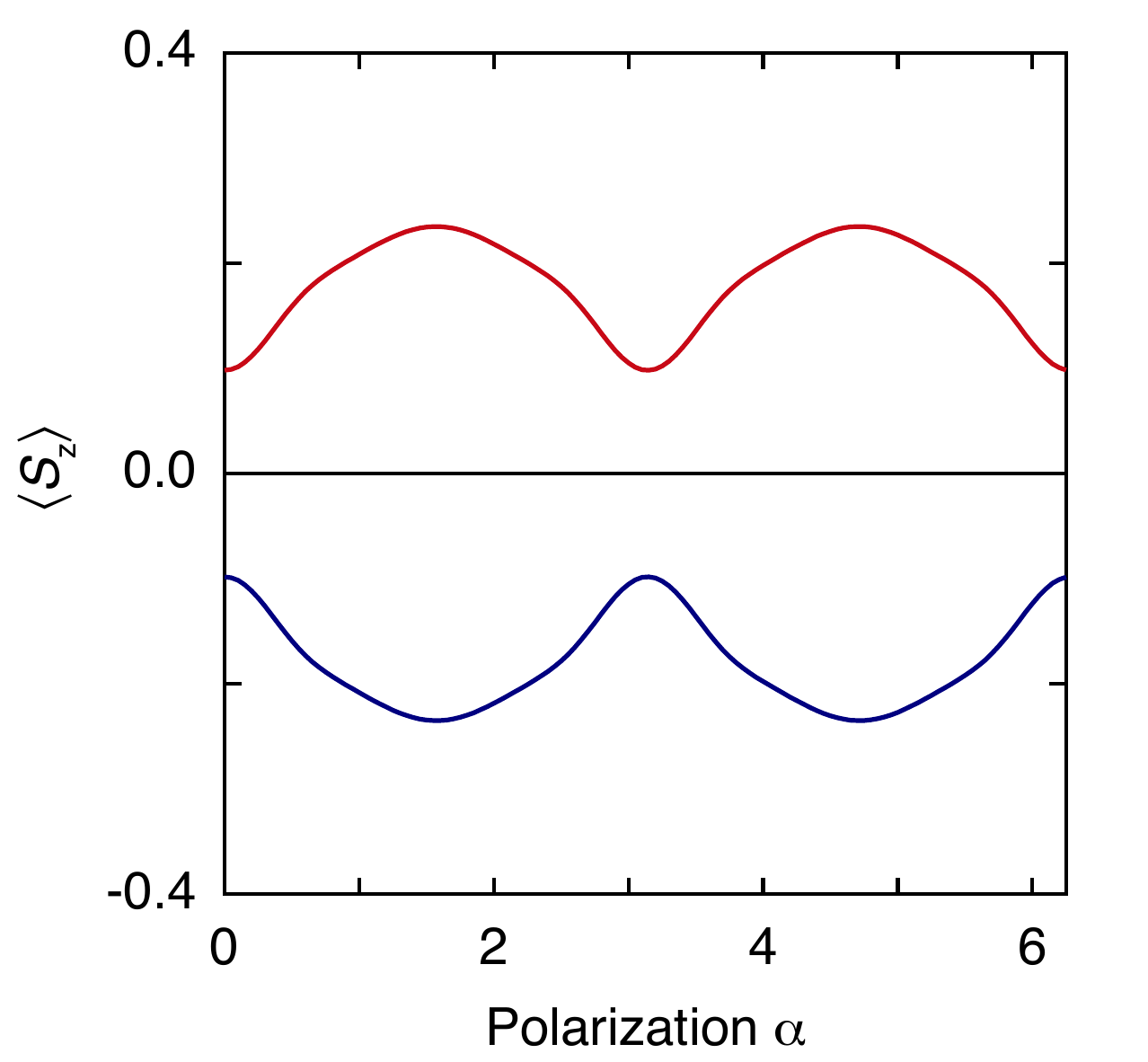}
\caption{Staggered $S^z$ with $L=((1.4,1),(0.2,-0.2),(1,1.4))$, with $C_3$-breaking and $C_2$-symmetry-allowed dissipation, at long times (steady state, $t = 1000$) as a function of the polarization angle $\alpha$ in $\mathcal{E}_x = \cos \alpha$ and $\mathcal{E}_y = \sin \alpha$, and $\Gamma_2 = 1$ (working in units of $\Gamma_1 = 1$).}
\label{fig:FigS13}
\end{center}
\end{figure}

The $\langle S^z \rangle$ expectation values are shown as a function of the polarization angle $\alpha$ in Fig.~\ref{fig:FigS12}. We notice a clear $2 \alpha$-dependence with nodal points. This does not yet match the experimental observation that on resonance the signal is mostly independent of $\alpha$. This leads us to hypothesize that the assumed decay channels as derived are too strongly constrained by the imposed symmetries. To check this, we use a more general jump operator (that preserves $C_2$ but no longer $C_3$), which might be due to the $C_3$ breaking layer stacking. The resulting angular dependence is shown in Fig.~\ref{fig:FigS13}. We see that there is some $\alpha$-independent contribution, with some weak angular dependence on top of this background, in good agreement with the experimental results.

\bibliographystyle{apsrev4-2}

\end{document}